\documentclass[lettersize,journal]{IEEEtran}
\usepackage{amsmath,amsfonts}
\usepackage{algorithmic}
\usepackage{algorithm}
\usepackage{array}
\usepackage{textcomp}
\usepackage{stfloats}
\usepackage{url}
\usepackage{verbatim}
\usepackage{graphicx}
\usepackage{cite}
\usepackage{multirow}
\usepackage{makecell}
\usepackage{bbding}
\usepackage{utfsym}
\usepackage{pifont}
\usepackage{hyperref}
\usepackage{subfigure}
\usepackage{amssymb}
\usepackage{booktabs}
\usepackage[normalem]{ulem}

\begin{document}

\title{ROSE: A Recognition-Oriented Speech Enhancement Framework in Air Traffic Control Using Multi-Objective Learning}

\author{Xincheng Yu, Dongyue Guo, Jianwei Zhang, and Yi Lin,~\IEEEmembership{Senior Member, IEEE}
\thanks{This work was supported by the National Natural Science Foundation of China (NSFC) under grants No. 62371323 and U2333209. (Corresponding author: Yi Lin)}
\thanks{X. Yu, D. Guo, J. Zhang, and Y. Lin are with the College of Computer Science, Sichuan University, Chengdu 610000, China (e-mail:  yuxincheng1998@stu.scu.edu.cn, dongyueguo@stu.scu.edu.cn, zhangjianwei@scu.edu.cn, yilin@scu.edu.cn)}}

\markboth{Journal of \LaTeX\ Class Files,~Vol.~14, No.~8, August~2021}%
{Shell \MakeLowercase{\textit{et al.}}: A Sample Article Using IEEEtran.cls for IEEE Journals}

\IEEEpubid{0000--0000/00\$00.00~\copyright~2021 IEEE}

\maketitle

\begin{abstract}
    Radio speech echo is a specific phenomenon in the air traffic control (ATC) domain, which degrades speech quality and further impacts automatic speech recognition (ASR) accuracy. In this work, a time-domain recognition-oriented speech enhancement (ROSE) framework is proposed to improve speech intelligibility and also advance ASR accuracy based on convolutional encoder-decoder-based U-Net framework, which serves as a plug-and-play tool in ATC scenarios and does not require additional retraining of the ASR model. Specifically, 1) In the U-Net architecture, an attention-based skip-fusion (ABSF) module is applied to mine shared features from encoders using an attention mask, which enables the model to effectively fuse the hierarchical features. 2) A channel and sequence attention (CSAtt) module is innovatively designed to guide the model to focus on informative features in dual parallel attention paths, aiming to enhance the effective representations and suppress the interference noises. 3) Based on the handcrafted features, ASR-oriented optimization targets are designed to improve recognition performance in the ATC environment by learning robust feature representations. By incorporating both the SE-oriented and ASR-oriented losses, ROSE is implemented in a multi-objective learning manner by optimizing shared representations across the two task objectives. The experimental results show that the ROSE significantly outperforms other state-of-the-art methods for both the SE and ASR tasks, in which all the proposed improvements are confirmed by designed experiments. In addition, the proposed approach can contribute to the desired performance improvements on public datasets.
\end{abstract}


\section{Introduction}
\label{sc:introduction}
\IEEEPARstart{I}{n} air traffic control (ATC), speech is a primary means to achieve air-ground communication between the air traffic controller (ATCO) and the aircrew. ATCO issues spoken ATC instructions via the very high frequency radio to provide required services for aircrew, whereas the aircrew reads back the received instructions promptly to ensure a consentient understanding of the flight operation. In this procedure, the automatic speech recognition (ASR) technique is a preferred solution to convert the spoken instruction into readable texts, based on which contextual ATC elements are extracted to obtain traffic dynamics and further support ATC operations \cite{RN198, RN181}.

Recently, enormous efforts have been made to achieve the ASR task \cite{RN182,RN197,RN196}. By analyzing the ASR results, volatile noise is a major issue that limits ASR performance in the ATC environment, referring to the environment noise, device noise, radio transmission noise, speech echo, etc. Unlike the additive non-stationary signal (i.e., environment, device, radio), the speech echo is a specific overlapping phenomenon dedicatedly generated by the ATC communication between the sent and received ATCO speech \cite{WOS:001058793500002}.

\begin{figure}[t!]
\centering
\subfigure[]{
    \begin{minipage}[t]{0.45\linewidth}
        \centering
        \includegraphics[width = \linewidth]{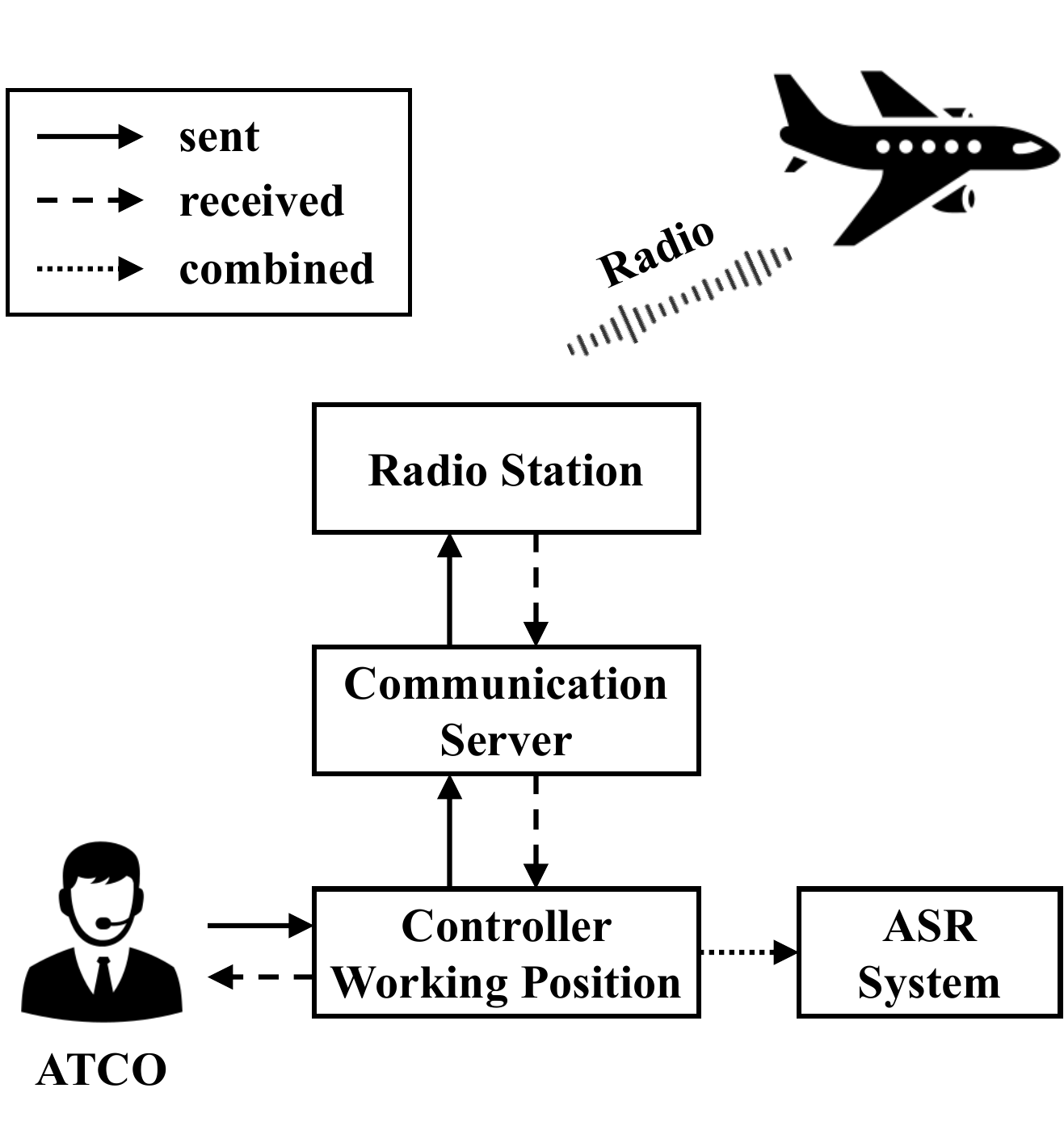}
    \end{minipage}%
}%
\subfigure[]{
    \begin{minipage}[t]{0.5\linewidth}
        \centering
        \includegraphics[width = \linewidth]{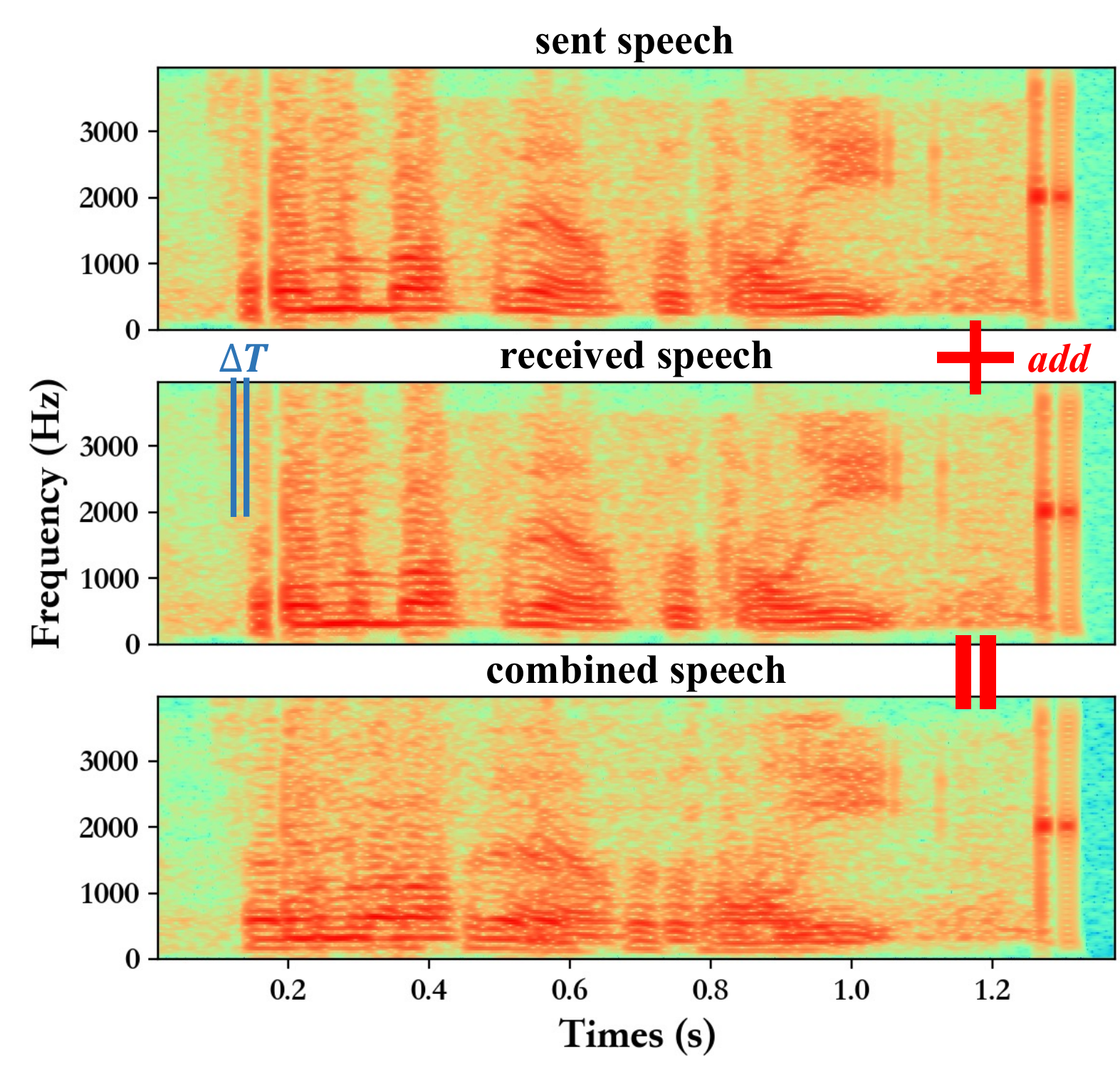}
    \end{minipage}%
}%
\caption{(a) The mechanism of the ATC speech transmission. (b) The speech spectrograms of different procedures generated by the CWP.}
\label{fig:background}
\end{figure}

\IEEEpubidadjcol
As shown in Figure \ref{fig:background} (a), the spoken instructions issued by ATCO are passed to the controller working position (CWP), the communication server, and finally to the radio stations, which are further forwarded to the aircrew (sent speeches). To ensure that the issued ATCO spoken instructions can be successfully received by receivers, based on the ATC procedure, the sent ATCO instructions are also returned to the assistant ATCO of the corresponding CWP to allow the ATCO to confirm the communication reliability (received speeches). Finally, the combined speeches are determined by addition operation based on the sent and received speeches in the CWP and fed into the following ASR systems, which can provide an integrated speech interface and reduce the system complexity. 

Due to the returned ATCO instructions, a temporal delay between the sent speech and the received speech exists in the combined speech, which is defined as \emph{\textbf{speech echo}}. As shown in Figure \ref{fig:background} (b), the speech (only from ATCO) collected from CWP is outlined in the spectrogram manner and the time offset ($\Delta$T) is presented in the received speech. Compared to uncombined signals (sent or received speech), the combined speech spectrogram is distorted by redundant echo interference of signal overlapping, which further degrades the perception and intelligibility of the ATCO speech.

By investigating the ASR performance, the accuracy of the ATCO speech (with speech echo) is obviously lower than that of the aircrew speech (without speech echo). Therefore, compared to other additive non-stationary noises, due to the ATC specificities, it is believed that the ASR performance can be significantly advanced by tackling the speech echo to enhance the speech quality.

Since the temporal delay is affected by many factors (distance between CWP and radio station, radio frequency, etc.), the value of $\Delta$T is varied between 10 and 200 milliseconds, which is distinct for each utterance. Note that both the sent and received speech that comprise each combined speech are issued by the same ATCO and have the same text. The term “echo" describes the operation of a radio station returning a received speech with the same text as the sent speech. Although the type of signal that leads to the generation of the speech echo cannot be precisely defined, it can be treated as special coupling noise and is not the solo reverberation or echo even though it is called “echo" in the ATC domain.

As the front-end module of the ASR system, speech enhancement (SE) is applied to eliminate noise interference on speech signals to obtain clean speech with higher intelligibility. The enhanced clean speeches, i.e., the SE output, will subsequently serve as the input of the ASR system. In the rest of this paper, the ATCO speeches with echo are regarded as noisy speeches, while their corresponding echo-free counterparts are clean speeches.

However, previous works \cite{loizou2010reasons, wang2019bridging} found that the enhanced signal from the SE model might yield inferior recognition accuracy for the subsequent ASR task. Some vital latent ASR-related information in the original noisy signal is diminished by the SE procedure with the redundant noise due to the different optimization targets between the SE and ASR tasks. 

The strategy of joint optimization is an effective means. Some studies \cite{pandey2021dual, ma2021multitask} proposed a joint training approach to optimize the SE and ASR models together using enhanced signal only. Recent work \cite{hu2022interactive} further combined enhanced and noisy signal features by utilizing an interactive fusion network to compensate for the neglected information in the original signal. 

Although these methods have achieved good performance on SE and ASR tasks, they inevitably involve parameter modification of the ASR module. In the ATC domain, existing ASR models are trained with thousands of hours of speech corpora and have been deployed in over 20 real-world industrial ATC environments \cite{RN196}. Therefore, retraining and redeploying the ASR models requires considering many issues, including data collecting, data annotation, extra computational resources and deployment costs. Moreover, the acquisition of high-quality speech is necessary to further support auditory-related downstream tasks in ATC scenarios such as emotion recognition and pilot/ATCO role identification.

To address this issue, in this work, a novel recognition-oriented speech enhancement (ROSE) framework is proposed to improve both speech intelligibility and ASR performance, which serves as a plug-and-play tool in ATC scenarios and does not require additional retraining of the ASR model. Inspired by previous works \cite{RN148,kim2021se}, ROSE is implemented using an encoder-decoder-based U-Net architecture with skip connections in the time domain. To enhance the model capacity for a multi-dimensional and robust feature space, two attention modules are designed to optimize the extracted feature representations. To achieve the recognition-oriented SE model, ASR-oriented optimization targets are designed to improve recognition performance in the ATC environment by learning robust feature representations, in which the multi-objective learning mechanism is applied to achieve joint optimization of SE and ASR objectives.

Specifically, the SE is to generate similar time and frequency domain features for the noisy speeches by referring to their clean counterparts. Based on the previous works \cite{RN148,kim2021se}, the mean absolute error (MAE) loss in the time domain and log-scale short-time Fourier transform (STFT)-magnitude loss \cite{RN188} in the frequency domain serve as the optimization targets for the SE task.

In addition, for a certain ASR model, the enhanced output of the SE model for noisy signals only obtains the desired performance by generating similar ASR-related features to that of clean signals. To this end, the well-designed algorithms for handcrafted feature engineering, such as spectrogram, Mel-frequency cepstral coefficients (MFCC), etc., inspire us to construct the optimization targets to retain the ASR performance for the SE output based on Fourier-based features.

In the encoder-decoder-based U-Net architecture, skip connections are built to generate fused features between encoder layers and corresponding decoder layers. To enhance the feature fusion, an attention-based skip-fusion (ABSF) module is applied to mine shared features from encoders using an attention mask, which prevents the model from concatenating the hierarchical features directly. 

To enhance the feature spaces, a channel and sequence attention (CSAtt) module is designed to learn objective-oriented feature weights in dual parallel attention paths. The channel attention path focuses on the feature dimension to excite the informative points and suppress disturbance features. The sequence attention path is expected to allocate dedicated weights to highlight salient frames and discard echo frames, which further enhances the feature discriminative ability.

To advance the proposed SE model into ATC practices, a real-world monaural speech corpus collected from ATC scenarios is used to validate the proposed approach considering the dedicated generation mechanism of the ATC communication. The experimental results demonstrate that the proposed framework outperforms other comparative baselines, obtaining considerable improvements on both subjective and objective evaluation metrics of SE task.  Most importantly, the enhanced speeches of the proposed model can achieve comparable ASR performance to clean speeches without any model retraining, i.e., 3.50\% character error rate (CER) on the selected ASR model. In summary, this work contributes to the SE study in the ATC domain in the following ways:

1) To the best of our knowledge, this is the first work that applies the SE algorithm to address the speech echo in the ATC environment. A novel end-to-end deep architecture ROSE is proposed to achieve the SE and ASR tasks in the time domain, which obtains high-quality spoken signal and achieves the desired ASR accuracy without retraining the existing models. The framework can also be directly applied in a real-world ATC environment.

2) In the proposed model, two attention modules are designed to optimize the extracted feature representations. An ABSF module is applied to skip connections to achieve feature fusion using an attention mask. A CSAtt module is innovatively designed to learn informative features and suppress disturbance factors from channel and sequence dimensions.

3) To improve recognition performance in the ATC environment, ASR-oriented objectives are designed to learn robust ASR features by optimizing the spectral distance between noisy and clean speeches. Moreover, a multi-objective learning mechanism is formulated to optimize the ROSE model using both the SE-related and ASR-related loss functions.

4) Extensive experimental results demonstrate that the ROSE outperforms other baselines on both SE and ASR tasks, and all the proposed technical improvements make expected contributions to related tasks. Most importantly, the proposed ROSE can be generalized to public open-source datasets and obtain comparable performance with state-of-the-art (SOTA) models.

\section{Related Works}
\subsection{The ATC-Related ASR Works}
In the ATC domain, the ASR technique aims at converting spoken ATC instructions into structured ATC-related elements \cite{RN194}, which is regarded as a key procedure to enhance the efficiency and safety of traffic operation \cite{RN195}. Regarding multilingual ASR in the ATC scenarios (i.e., Chinese and English), a unified framework was constructed based on the acoustic model, pronunciation model and language model\cite{RN181, RN196}. The feature representation was studied to enhance the ASR performance \cite{RN186}. Semi-supervised training strategy was presented to address the small sample problem in the ATC domain \cite{RN184}. The transfer learning was introduced to enhance the model applicability for different scenes \cite{RN197}. The contextual information of the real-time traffic operation was fused to improve the accuracy of ATC elements \cite{RN198, RN199}. 

The ASR technique is widely applied to enhance ATC applications, such as monitoring flight safety \cite{RN195}, reducing the controller’s workload \cite{RN200}, the ATCO training system \cite{RN201} and decision support \cite{RN202}. According to \cite{RN196}, it is believed that enhancing speech quality is a feasible solution to improve ASR performance, which is the primary focus of this work.

\subsection{The SE Works}
In the past few decades, SE has been successfully applied to hearing aids \cite{RN203}, audio and video calls \cite{RN204}, voice conversion \cite{RN205}, speaker identification, and ASR system \cite{RN206, RN207}.

Traditional SE methods \cite{RN162,RN102} usually use heuristic or direct algorithms to achieve denoising tasks. With the development of deep learning techniques, deep neural networks (DNN) are introduced into SE research, which have shown potent noise reduction ability in complex situations.

DNN-based SE models are divided into time-frequency (T-F) and time-domain methods. T-F domain methods convert the raw signal to the corresponding spectral magnitude as input by using STFT processing, which have achieved the desired performance on SE task \cite{fu2019metricgan,meganplus,Yu2023HighFS}. Although the magnitude represents T-F information of the signal, some limitations are existing in the T-F methods \cite{wang2021tstnn}. Existing study \cite{takahashi2018phasenet} has proven that phase information is also crucial for enhancing speech quality. Recent works consider complex features and combine them with spectral magnitude or phase information to further achieve better performance \cite{dccrn,yin2020phasen,yu2022dual,wang2023tf}.

Time-domain methods directly enhance noisy speech from raw waveform in an end-to-end manner. Various works have been achieved in convolutional neural networks (CNN), recurrent neural networks and generative adversarial networks (GAN) \cite{RN100, RN148,wang2021tstnn,shin2022multi,kim2021se}. Since the raw waveform contains the full information of the signal, the paradigm of learning directly from the waveform avoids the need for intermediate transformations (STFT and inverse STFT) and fully leverages the potentially valuable information (e.g., phase). Therefore, we propose the time-domain-based SE method to achieve the denoising task in the ATC domain.

\begin{figure*}[htbp]
    \centering
    \includegraphics[width=0.9\linewidth]{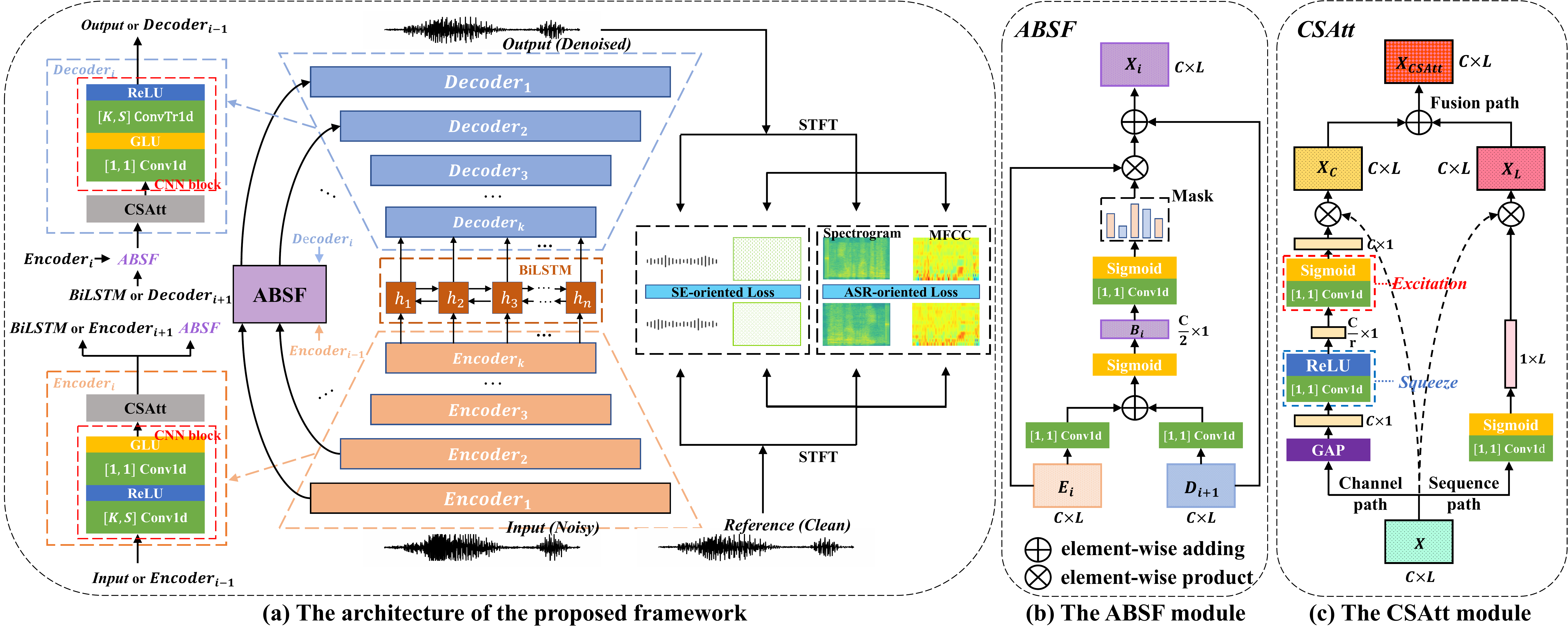}
    \caption{Overall network architecture of ROSE and details of the ABSF module and CSAtt module.}
    \label{pic:ROSE}
\end{figure*}

\section{ROSE}
\subsection{Problem Formulation}
In general, the supervised SE task can be regarded as denoising the noisy speech signal to obtain an enhanced signal that is close to the reference signal. In the ATC domain, given the noisy combined speech with the speech echo $x=\{x_{1}, x_{2}, ..., x_{T}\}$ composed of the received speech $s=\{s_{1}, s_{2}, ..., s_{T}\}$, the sent speech with $\Delta T$ temporal delay $s^{'}=\{s^{'}_{1}, s^{'}_{2}, ..., s^{'}_{T}\}$ and additional intrusive noise $n=\{n_{1}, n_{2}, ..., n_{T}\}$ such that $x=s+s^{'}+n$, the goal of the end-to-end SE task is to obtain the denoised signal $\tilde{s}$ directly from the noisy signal $x$, as Equation \ref{eq:fx}:
\begin{equation}
    \label{eq:fx}
    \tilde{s} = \{\tilde{s}_{1}, \tilde{s}_{2}, ..., \tilde{s}_{T}\} = f(x)
\end{equation}
where $T$ is the number of signal samples and $f(\cdot)$ denotes the SE model.

\subsection{Overview}
As shown in Figure \ref{pic:ROSE} (a), the proposed framework is implemented using an end-to-end deep architecture, in which the backbone consists of a multi-layer encoder-decoder structure with U-Net skip connections by referring to \cite{RN148}. In this work, the index of the decoder layer is the same as that of the corresponding encoder layer for convenient descriptions. The bidirectional long short-term memory (LSTM) is designed to capture the temporal dependencies among the frames and mine the correlations of signals in the past and future dimensions. 

The basic encoder block is comprised of sequential CNN blocks (1-d convolution (Conv1d) with kernel size $K$ and stride $S$, rectified linear unit (ReLU), pointwise Conv1d with kernel size 1 and stride 1, gated linear unit (GLU)) and CSAtt modules. To formulate a U-Net shape, the basic decoder block applies an architecture similar to the encoder block, but with a different stacking order, i.e. CSAtt modules and (pointwise Conv1d, GLU, transposed 1-d convolution (ConvTr1d), ReLU). The ABSF modules are integrated into the skip connections to achieve recalibration of features that are transmitted from encoder layers.

\subsection{Attention-based Skip-fusion}
\label{sc:absf}
In general, skip-connection is to transfer shallow encoder features into decoder blocks to formulate robust features. However, current practices fuse the encoder and decoder features by addition or concatenation operations with the same weight, which may not be optimal for complex tasks \cite{RN100,RN148}. In this study, the ABSF module is designed to optimize the learned features, in which the gate block is to extract involved correlations from encoded features by multiplying them with an attention mask.

As shown in Figure \ref{pic:ROSE} (b), let the learned feature of the $i$-th encoder layer be $E_i\in \mathbb{R}^{C \times L}$, and the learned feature of the $(i+1)$-th decoder layer be $D_{i+1}\in \mathbb{R}^{C \times L}$. As shown in Equation \ref{eq:bi}, the 1-d convolution layers are applied to the above two feature maps to halve the number of channels, and the concatenation operation is performed to initialize the fusion attention coefficient matrix $B_i$. 
\begin{equation}
    \label{eq:bi}
    B_{i}=\sigma\left(\psi_{\frac{C}{2}}^{[1, 1]}\left(E_{i}\right) \oplus \psi_{\frac{C}{2}}^{[1, 1]}\left(D_{i+1}\right)\right), B_{i} \in \mathbb{R}^{\frac{C}{2} \times L}
\end{equation}
where $\psi_{C}^{[K, S]}$ are convolution operations with kernel size $K$ and stride $S$ of $C$ filters, $\oplus$ denotes the element-wise adding operation and $\sigma$ denotes the sigmoid activation. 

Based on the $B_i$, a 1-d convolution layer followed by a sigmoid activation is used to project the number of channels as the same as the input shape, further obtaining an attention mask $A_i$. Finally, the output $X_i$ is obtained using the following rules, in which $\otimes$ denotes element-wise product operation. 
\begin{equation}
A_{i}=\sigma\left(\psi_{C}^{[1, 1]}\left(B_{i}\right)\right), A_{i} \in \mathbb{R}^{C \times L}
\end{equation}
\begin{equation}
X_{i}=D_{i+1} \oplus\left(E_{i} \otimes A_{i}\right), X_{i} \in \mathbb{R}^{C \times L}
\end{equation}

\subsection{The Channel and Sequence Attention}
\label{sc:csatt}
Facing the temporal shift of the speech echo, the transitions between neighbor phonemes are overlapped to cause disordered representations, further impacting the ASR performance. In this work, a dual-attention module is proposed to recalibrate the learned features in a parallel manner, which assigns task-oriented weights on both channel and sequence dimensions, i.e., CSAtt. The proposed module achieves dual attention, which is illustrated as follows:

\textbf{Channel Path.} This path aims to reinforce the model to perceive informative feature channels using the attention mechanism. Considering the fact that the noise and spoken words are distributed in different frequencies of the speech features, channel attention is desired to emphasize the features with higher weights for spoken words and suppress features with lower weights for the noise.

\textbf{Sequence Path.} This path is to guide the model to focus on selective frames by employing higher attention weights. Specifically, the trainable attention maps focus on the marginal frames to remove echo features to support the SE task and also highlight the central frames of each phoneme to advance the ASR task. 

\textbf{Fusion path.} This path is to fuse features learned from channel and sequence paths and feed them into subsequent modules.

The global average pooling (GAP) operation is applied to generate the sample-related initial weights based on the input feature map. As shown in Figure \ref{pic:ROSE} (c), let the input feature map be $X\in \mathbb{R}^{C \times L}$, where $C$ is the number of channels, $L$ is the length of the sequence dimension. The GAP operation is applied to generate the sample-related initial weights based on the input feature map along the channel dimension.
The squeeze-and-excitation attention \cite{hu2018squeeze} is to obtain the channel attention weights $W_C$, as shown below:
\begin{equation}
\label{eq:xb}
X_{B}=\delta\left(\psi_{\frac{C}{r}}^{[1, 1]}(GAP(X))\right), X_{B} \in \mathbb{R}^{\frac{C}{r} \times 1}
\end{equation}
\begin{equation}
\label{eq:wc}
W_{C}=\sigma\left(\psi_{C}^{[1, 1]}(X_{B})\right), W_{C} \in \mathbb{R}^{C \times 1}
\end{equation}
where Equations \ref{eq:xb} and \ref{eq:wc} denote the channel squeeze and excitation operations, in which the 1-d convolution layer followed by the ReLU activation is used to squeeze the number of channels and the 1-d convolution layer followed by the sigmoid activation is used to excite the number of channels, respectively. $r$ is a hyper-parameter to indicate the squeezed degree. $\delta$ denotes the ReLU activation.

Finally, the output feature of the channel attention $X_C$ is obtained by the following rules: 
\begin{equation}
    X_{C}=X\otimes W_{C}, X_{C}\in \mathbb{R}^{C \times L}
\end{equation}

Similarly, a 1-d convolution layer followed by a sigmoid activation is applied to reduce the number of channels to 1, obtaining the sequence attention weights $W_L$. The sequence-attentive feature $X_L$ is obtained by Equation \ref{eq:xl}:
\begin{equation}
W_{L}=\sigma\left(\psi_{1}^{[1, 1]}(X)\right), W_{L} \in \mathbb{R}^{1 \times L}
\end{equation}
\begin{equation}
\label{eq:xl}
X_{L}=X\otimes W_{L}, X_{L} \in \mathbb{R}^{C \times L}
\end{equation}

Finally, The fused features of the CSAtt module are achieved as below: 
\begin{equation}
    X_{CSAtt}=X_{C}\oplus X_{L}, X_{CSAtt}\in \mathbb{R}^{C \times L}
\end{equation}

\subsection{Multi-objective Loss Function}
As illustrated before, the proposed model is optimized by multi-objective learning to enhance speech quality and ASR-required features. Based on previous works \cite{RN148,kim2021se}, the MAE loss in the time domain and log-scale STFT-magnitude loss in the frequency domain serve as the optimization targets for the SE task. 

Given $s$ and $\tilde{s}$ be the clean and denoised speech signal respectively, the loss function of the SE task $L_{SE}$ is obtained using the following rules:
\begin{equation}
L_{M A E}=\|s-\tilde{s}\|_{1}
\end{equation}
\begin{equation}
L_{M A G}=\left\|\log \left|S T F T\left(s ; \theta\right)\right|-\log \left|STFT\left(\tilde{s} ; \theta\right)\right|\right\|_{1}
\end{equation}
\begin{equation}
L_{S E}=L_{M A E}+L_{M A G}
\end{equation}
where $\left \|*\right \|_{1}$ denotes the $L_1$ norm. $\left|S T F T(*)\right|$ denotes the magnitude of signal processed through STFT. $\theta$ denotes the STFT parameters (FFT bins, hop sizes, window lengths).

For the ASR task, the spectral convergence (SC) loss is regarded as the optimization target to achieve ASR-required features. According to \cite{RN190}, the core idea is to optimize the spectral distance between the enhanced speech and the reference speech, aiming at guiding the model to capture the informative ASR-related information embedded in the signal spectrums. The SC loss can be defined as follows:
\begin{equation}
    f\left(D ; \theta\right)=\left\|D\left(s ; \theta\right)-D\left(\tilde{s} ; \theta\right)\right\|_{F}/\left\|D\left(s ; \theta\right)\right\|_{F}
\end{equation}
where $\left \|*\right \|_{F}$ denotes the Frobenius norm and $D$ denotes the handcrafted features of the speech signal. In this work, as the input features of the existing ASR models in the ATC domain \cite{RN181,RN194,RN197}, spectrogram and MFCC features are selected to achieve ASR-oriented loss $L_{ASR}$, as shown below:
\begin{equation}
L_{Spec. } = f\left( \text{Spectrogram}; \theta\right),\hspace{0.2cm}
L_{M F C C} = f\left(\text{MFCC} ; \theta\right)
\end{equation}
\begin{equation}
L_{A S R} = L_{Spec. }+L_{M F C C}
\end{equation}

Finally, the total loss is formulated to achieve the multi-objective training with two hyper-parameters as below:
\begin{equation} 
\label{eq:multi}
    L=\lambda_1 L_{SE}+\lambda_2 L_{ASR}
\end{equation}

\section{Experiments}
\subsection{Dataset}
In this work, three datasets collected from real-world industrial ATC systems are used to validate the proposed approach, including clean-noisy pair dataset \cite{WOS:001058793500002}, ATCSpeech-large dataset \cite{RN186} and SSL dataset \cite{guo2023boosting}. Specifically, The clean-noisy pair dataset contains only ATCO's speech data and is used to train and test proposed ROSE and other baseline methods. The detailed descriptions of dataset division are shown in Table \ref{tab:tabledataset} and more relevant details can be found in \cite{WOS:001058793500002}. In addition, the text labels are annotated for the dataset to evaluate the ASR performance, including a total of 683 Chinese characters and 437 English words (a total of 712 tokens in the vocabulary, including 683 Chinese characters, 26 English letters and 3 special tokens, i.e., SPACE, UNK and '). The ATCspeech-large dataset contains a total of 1275 hours of annotated speech data, including 701 Chinese characters and 1221 English words (a total of 730 tokens in the vocabulary). The SSL dataset contains two sub-datasets for pre-training (SSL) and fine-tuning (SL) processes, respectively, in which the SSL set contains a total of 1032 hours of unlabeled speech data and the SL set contains a total of 15 hours of annotated speech data, including a total of 639 Chinese characters and 584 English words (a total of 668 tokens in the vocabulary). All datasets are independent and the latter two datasets are used separately to train the two selected ASR models.

\begin{table}[!t]
  \caption{Data size of the dataset, \#U denotes the speech utterances and \#H denotes the speech durations (Hours).\label{tab:tabledataset}}
  \centering
  \begin{tabular}{lcccccc}  
  \hline  
  \cline{1-7}
  \multicolumn{1}{c}{\multirow{2}*{Language}} & \multicolumn{2}{c}{Train} & \multicolumn{2}{c}{Validation} & \multicolumn{2}{c}{Test} \\ \cline{2-7}
  \multicolumn{1}{c}{}                          & \#U         & \#H         & \#U        & \#H        & \#U         & \#H        \\ \hline
  Chinese                                       & 42189       & 37.28       & 4188       & 3.69       & 6012        & 5.08       \\
  English                                       & 5064        & 5.55        & 558        & 0.62       & 502         & 0.54       \\ \hline
  Total                                         & 47253       & 42.83       & 4746       & 4.31       & 6514        & 5.62       \\ \cline{1-7}   \hline
  \end{tabular}
  \end{table}

\subsection{Experimental Configurations}
\label{sc:Configurations}
In the ROSE, the depth of both the encoder and decoder are 5 and the hidden dimension of each layer is $H=48$. For the basic CNN block of encoder/decoder, Conv1d or ConvTr1d is configured with kernel size $K=8$ and stride $S=4$. For the CSAtt module, $r$ is set to 2 for squeezing the feature channels. Two bidirectional LSTM layers with 768 neurons are designed to capture the sequential dependencies. For the STFT parameters, FFT bins are 512, hop sizes are 100, and window lengths are 400. The dimension of MFCC is set to 13. The $\lambda_{i}$ for the loss function in Equation \ref{eq:multi} is set to 1. The Adam optimizer is applied to optimize the proposed framework with an initial learning rate of 3e-4 and 0.999 decay. All samples in the clean-noisy pair dataset are upsampled from 8 kHz to 16 kHz and clipped to 4 seconds long for fair comparisons with baseline models. The batch size is set to 64 during the model training.

To confirm the effectiveness and efficiency of the proposed model on ASR performance, two optimized ASR models are applied to evaluate the clean-noisy pair test set, including a supervised Deepspeech 2 (DS2) model \cite{RN209} and a self-supervised learning (SSL)-based Wav2Vec 2.0 (W2V2) model \cite{guo2023boosting}. Specifically, \emph{1:)} the ATCSpeech-large dataset is applied to train the DS2 model, in which the input feature is the 81-dimension spectrogram. The Adam optimizer is applied to optimize the model parameters with an initial learning rate of 1e-3 and 0.99 decay. \emph{2:)} The SSL set is applied to pre-train the W2V2 encoder to obtain a pre-trained model, and the SL set is applied to supervisedly fine-tune the pre-trained W2V2 model to obtain an ASR-focused fine-tuned model. The W2V2 encoder is optimized with Adam and trained for 450k updates, warming up the learning rate for the first 8\% of updates to a peak of 5 $\times 10^{-4}$, and then linearly decaying it. The finetuning process is implemented on fairseq\footnote[1]{\url{https://github.com/pytorch/fairseq/tree/master/examples/wav2vec}} with the Connectionist Temporal Classification (CTC) loss function. All the above ASR models are end-to-end multilingual models without language models. Note that the ASR experiments designed in this work focus on validating the proposed ROSE on both SE and ASR tasks in the ATC domain.

\begin{table}[!t]
    \caption{Details of the AIR subjective rating.\label{tab:tableatc}}
    \centering
    \begin{tabular}{cc}
    \hline
    \multicolumn{1}{c|}{Score} & Descriptions                         \\ \cline{1-2}
    \multicolumn{1}{c|}{1}     & Unable to understand spoken words     \\ \cline{1-2}
    \multicolumn{1}{c|}{2}     & \makecell[c]{Some spoken words can be understood, but \\ cannot be organized as ATC instructions}        \\ \cline{1-2}
    \multicolumn{1}{c|}{3}     & \makecell[c]{Can understand most spoken words of \\the ATC instructions}  \\ \cline{1-2}
    \multicolumn{1}{c|}{4}     & \makecell[c]{Can understand ATC instructions clearly and \\repeat the spoken words}         \\ \hline
    \end{tabular}
\end{table}

\subsection{Baseline Methods}
\label{sc:baselinemodel}
In this work, several open-source SOTA baselines with different types are selected to validate the proposed model, which are classified as follows: \emph{1) traditional methods:} Wiener \cite{RN162} and spectral subtraction (Spec-Sub)\cite{RN102}, \emph{2) GAN-based generative methods:} SEGAN \cite{RN100}, MetricGAN \cite{fu2019metricgan} and MetricGAN+ \cite{meganplus}, \emph{3) Diffusion-based generative methods:} DiffuSE \cite{diffuse}, CDiffuSE \cite{cdiffuse} and SGMSE \cite{sgmse} and \emph{4) discriminative methods:} CRN \cite{crn}, DCCRN-E \cite{dccrn}, DEMUCS \cite{RN148}, BSRNN \cite{Yu2023HighFS}, SE-Conformer \cite{kim2021se}, FullSubNet+ \cite{fullsubnet}, DB-AIAT \cite{yu2022dual} and TF-GridNet \cite{wang2023tf}. These methods cover time (T) and T-F domain models, as well as causal and non-causal paradigms. All of those methods are conducted on the official configurations and adapted to the ATC corpus following the relevant works.

\subsection{Evaluation Metrics}
For the metrics of the SE task, we compute the following five typical objective measures on the test set, i.e., mean opinion score (MOS) prediction of the $i)$ signal distortion attending only to the speech signal (CSIG), $ii)$ intrusiveness of background noise (CBAK), $iii)$ the overall effect (COVL) (from 1 to 5) \cite{RN211}, perceptual evaluation of speech quality (PESQ) (from -0.5 to 4.5) \cite{RN212}, short-time objective intelligibility (STOI) (from 0 to 1) \cite{RN213}.

For the subjective evaluation, we randomly sample 10 utterances with complete ATC instructions and each one was scored by 45 ATC experts along two facets: \emph{1)} a relevant subjective MOS study is recommended in ITU-T P.835 \cite{2003Subjective}, including level of signal distortion (SIG), intrusiveness of background noise (BAK), and overall quality (OVL). \emph{2)} an ATC-related subjective measure is proposed to rate the ATC instructions recognition (AIR) and detailed descriptions are shown in Table \ref{tab:tableatc}. For all the above metrics, a larger score indicates a higher speech quality.

For the metric of the ASR task, the CER based on the Chinese character and English letter is applied to evaluate the ASR performance. We report the mean results and their 95\% confidence interval. For each model, the number of model parameters is reported in millions (M). We use the \emph{calflops\footnote[2]{\url{https://github.com/MrYxJ/calculate-flops.pytorch}}} toolkit to count the number of multiply-accumulate operations (MAC) for processing a 4-second signal, and report it in giga-operations per second (GMAC/s).

\begin{table*}[htbp]
    \caption{Objectvie experimental results for different methods on ATC corpus. The CER of the DS2 and W2V2 of the clean signal are \emph{(2.25$\pm$.16)\%} and \emph{(1.37$\pm$.15)\%}, respectively. \label{tab:ATCresults_1}}
    \centering
    \begin{tabular}{lcccc|ccccc|cc}
    \toprule
    \multirow{2}{*}{Model}    & \multirow{2}{*}{Cau.} & \multirow{2}{*}{Dom.} & \#Param. &\multirow{2}{*}{GMAC/s} & \multirow{2}{*}{CSIG$\uparrow$}   &\multirow{2}{*}{CBAK$\uparrow$}     &\multirow{2}{*}{COVL$\uparrow$}   & \multirow{2}{*}{PESQ$\uparrow$}    & \multirow{2}{*}{STOI$\uparrow$}     & \multicolumn{2}{c}{CER (\%)$\downarrow$}         \\ 
    & & &(M) & & & & & & & DS2 & W2V2 \\ \midrule
    Noisy  & - & - & - & -    & 3.28$\pm$.01   & 1.81$\pm$.01  & 2.48$\pm$.01  & 1.80$\pm$.01  & 0.56$\pm$.00   & 4.66$\pm$.24  & 2.81$\pm$.17 \\ 
    \cmidrule{1-12}
    \multicolumn{12}{c}{Traditional Methods}  \\
    \cmidrule{1-12}
    Wiener\cite{RN162} & \usym{2613} & T & - & -     & 2.17$\pm$.04  & 1.41$\pm$.03  & 1.66$\pm$.04  & 1.49$\pm$.04  & 0.54$\pm$.01   & 10.62$\pm$.42 & 7.12$\pm$.34\\
    Spec-Sub\cite{RN102} & \usym{2613} & T-F & - & - & 2.62$\pm$.04  & 1.63$\pm$.03  & 1.93$\pm$.03  & 1.67$\pm$.03  & 0.55$\pm$.00   & 9.44$\pm$.40 & 6.95$\pm$.31\\
    \cmidrule{1-12}
    \multicolumn{12}{c}{GAN-based Generative Methods}  \\
    \cmidrule{1-12}
    SEGAN\cite{RN100}     & \usym{2613} & T & 65.82 & 13.39  & 3.56$\pm$.04   & 2.55$\pm$.04  & 2.79$\pm$.04  & 2.12$\pm$.04  & 0.70$\pm$.01 & 9.74$\pm$.34 & 4.31$\pm$.27 \\
    MetricGAN\cite{fu2019metricgan} & \usym{2613} & T-F & 1.86 & 28.51 & 3.97$\pm$.04 & 2.96$\pm$.03  & 3.25$\pm$.03  & 2.70$\pm$.04  & 0.71$\pm$.00 & 8.31$\pm$.32 & 3.01$\pm$.24 \\
    MetricGAN+\cite{meganplus} & \usym{2613} & T-F & 1.86 & 28.51 & 4.13$\pm$.04 & 3.12$\pm$.03  & 3.41$\pm$.03  & 2.97$\pm$.04  & 0.72$\pm$.00 & 7.98$\pm$.30 & 2.89$\pm$.23 \\
    \cmidrule{1-12}
    \multicolumn{12}{c}{Diffusion-based Generative Methods}  \\
    \cmidrule{1-12}
    DiffuSE\cite{diffuse}  & \usym{2613} & T-F & 3.23 & 210.31 & 3.65$\pm$.04 & 2.63$\pm$.04  & 2.84$\pm$.03  & 2.21$\pm$.04  & 0.71$\pm$.00 & 11.54$\pm$.39 & 5.55$\pm$.29 \\
    CDiffuSE \cite{cdiffuse} & \usym{2613} & T-F & 4.28 & 190.23 & 3.70$\pm$.04 & 2.71$\pm$.03  & 2.90$\pm$.03  & 2.28$\pm$.04  & 0.71$\pm$.00 & 10.45$\pm$.37 & 5.02$\pm$.28 \\
    SGMSE \cite{sgmse} & \usym{2613} & T-F & 3.89 & 152.77 & 3.78$\pm$.03 & 2.80$\pm$.03  & 3.01$\pm$.04  & 2.39$\pm$.04  & 0.71$\pm$.00 & 9.44$\pm$.30 & 4.79$\pm$.25 \\
    \cmidrule{1-12}
    \multicolumn{12}{c}{Discriminative Methods}  \\
    \cmidrule{1-12}
    CRN \cite{crn}  & \Checkmark & T-F & 1.70 & 8.80 & 3.57$\pm$.04   & 2.51$\pm$.04  & 2.89$\pm$.04  & 2.33$\pm$.04  & 0.71$\pm$.00   & 7.97$\pm$.31 &2.91$\pm$.25 \\
    DCCRN-E\cite{dccrn} & \Checkmark & T-F & 3.67 & 33.48 & 3.76$\pm$.04  & 2.69$\pm$.03  & 3.07$\pm$.04  & 2.45$\pm$.03  & 0.71$\pm$.00 & 8.46$\pm$.34 &3.97$\pm$.29 \\
    DEMUCS \cite{RN148} & \Checkmark & T & 53.52 & 37.83 & 4.22$\pm$.04   & 2.91$\pm$.04  & 3.41$\pm$.04  & 2.62$\pm$.04  & 0.79$\pm$.01   & 4.89$\pm$.21  &2.75$\pm$.21  \\
    BSRNN \cite{Yu2023HighFS} & \Checkmark & T-F & 2.96 & 2.57 & 4.55$\pm$.04   & \textbf{3.26$\pm$.04}  & 3.85$\pm$.04  & 3.04$\pm$.04  & 0.83$\pm$.00   & 5.66$\pm$.24  &2.86$\pm$.29  \\
    DEMUCS \cite{RN148} & \usym{2613} & T & 68.87 & 38.12 & 4.25$\pm$.04   & 2.95$\pm$.04  & 3.45$\pm$.04  & 2.66$\pm$.04  & 0.80$\pm$.01   & 4.72$\pm$.20  & 2.72$\pm$.20  \\
    SE-Conformer\cite{kim2021se}  & \usym{2613} & T & 75.28 & 44.96 & 4.40$\pm$.03 & 3.15$\pm$.04  & 3.69$\pm$.04  & 3.02$\pm$.04  & 0.81$\pm$.00  & 4.28$\pm$.20 & 2.74$\pm$.19\\
    FullSubNet+\cite{fullsubnet}  & \usym{2613} & T-F & 8.67 & 48.57 & 4.12$\pm$.04 & 3.00$\pm$.04  & 3.45$\pm$.04  & 2.82$\pm$.03  & 0.75$\pm$.00 & 5.35$\pm$.24 & 2.76$\pm$.27 \\
    DB-AIAT \cite{yu2022dual} & \usym{2613} & T-F & 2.81 & 41.8 & 4.55$\pm$.05 & 3.25$\pm$.04  & \textbf{3.87$\pm$.04}  & 3.07$\pm$.04  & \textbf{0.84$\pm$.00} & 4.90$\pm$.24 & 2.79$\pm$.25\\
    TF-GridNet \cite{wang2023tf} & \usym{2613} & T-F & 8.2 & 19.2 & 4.53$\pm$.04 & 3.17$\pm$.04  & 3.80$\pm$.04  & 2.97$\pm$.04  & 0.82$\pm$.00 & 4.87$\pm$.23 & 2.83$\pm$.26\\
    \cmidrule{1-12}
    ROSE (Ours) & \usym{2613} & T & 36.98 & 4.73 & \textbf{4.58$\pm$.04} & 3.22$\pm$.04 & 3.85$\pm$.03 & \textbf{3.09$\pm$.03} & \textbf{0.84$\pm$.00} & \textbf{3.50$\pm$.18} & \textbf{2.42$\pm$.19}\\
    \bottomrule
    \end{tabular}
\end{table*}

\begin{table}[htbp]
    \caption{Subjective results for different methods on selected ATC samples. The selected DEMUCS is non-causal model. \label{tab:ATCresults_2}}
    \centering
    \begin{tabular}{l|cccc}
    \toprule
    \multirow{2}{*}{Model}    &MOS & MOS &MOS   & \multirow{2}{*}{AIR$\uparrow$}     \\ 
    &SIG$\uparrow$ &BAK$\uparrow$ &OVL$\uparrow$ &  \\ \midrule
    Noisy           & 3.42$\pm$.10  & 2.84$\pm$.09  & 2.91$\pm$.08  & 3.41$\pm$.03  \\
    \cmidrule{1-5}
    Wiener\cite{RN162}          & 2.78$\pm$.12	& 2.13$\pm$.11	& 2.26$\pm$.10  & 3.13$\pm$.04  \\ 
    Spec-Sub\cite{RN102}        & 2.98$\pm$.09 & 2.27$\pm$.10  & 2.39$\pm$.11  & 3.23$\pm$.03  \\
    \cmidrule{1-5}
    SEGAN\cite{RN100}           & 3.47$\pm$.10	& 2.85$\pm$.10  & 2.93$\pm$.09  & 3.43$\pm$.04  \\  
    MetricGAN\cite{fu2019metricgan}       & 3.76$\pm$.11  & 3.21$\pm$.09  & 3.03$\pm$.10  & 3.43$\pm$.03  \\ 
    MetricGAN+ \cite{meganplus}     & 3.78$\pm$.10  & 3.19$\pm$.09  & 3.02$\pm$.09  & 3.51$\pm$.04  \\ 
    \cmidrule{1-5}
    DiffuSE \cite{diffuse}        & 3.57$\pm$.10  & 2.84$\pm$.10  & 2.97$\pm$.10  & 3.41$\pm$.03  \\ 
    CDiffuSE \cite{cdiffuse}       & 3.55$\pm$.09  & 2.88$\pm$.10  & 2.96$\pm$.10  & 3.51$\pm$.03  \\ 
    SGMSE \cite{sgmse}          & 3.62$\pm$.10  & 2.97$\pm$.10  & 3.02$\pm$.08  & 3.62$\pm$.03  \\ 
    \cmidrule{1-5}
    CRN \cite{crn}            & 3.54$\pm$.08  & 2.78$\pm$.09  & 2.94$\pm$.10  & 3.47$\pm$.03  \\ 
    DCCRN-E \cite{dccrn}        & 3.62$\pm$.10  & 3.03$\pm$.10  & 2.95$\pm$.10  & 3.52$\pm$.03  \\ 
    DEMUCS \cite{RN148}& 3.89$\pm$.10	& 3.21$\pm$.10	& 3.15$\pm$.10  & 3.56$\pm$.04  \\ 
    BSRNN \cite{Yu2023HighFS}          & 3.85$\pm$.09  & 3.23$\pm$.10  & 3.13$\pm$.10  & 3.69$\pm$.03  \\ 
    SE-Conformer \cite{kim2021se}   & 3.83$\pm$.10  & 3.21$\pm$.10  & 3.13$\pm$.10  & 3.68$\pm$.03  \\ 
    FullSubNet+ \cite{fullsubnet}    & 3.82$\pm$.09  & 3.18$\pm$.08  & 3.10$\pm$.09  & 3.60$\pm$.03  \\ 
    DB-AIAT \cite{yu2022dual}        & 3.91$\pm$.10  & 3.20$\pm$.10  & \textbf{3.16$\pm$.09}  & 3.73$\pm$.03  \\ 
    TF-GridNet \cite{wang2023tf}     & 3.94$\pm$.10  & \textbf{3.25$\pm$.10}  & 3.10$\pm$.10 & 3.74$\pm$.04  \\
    \cmidrule{1-5} 
    ROSE (Ours)    & \textbf{3.98$\pm$.09}    & 3.24$\pm$.10	& 3.15$\pm$.10 & \textbf{3.78$\pm$.03}  \\ 
    \bottomrule
    \end{tabular}
\end{table}

\begin{table*}[htbp]
    \caption{The experimental results of the ablation study. \label{tab:ATCresults_ablation}}
    \centering
    \begin{tabular}{clcc|ccccc|cc}
    \toprule
    \multirow{2}{*}{Exp.} &\multirow{2}{*}{Model}    & \#Param. &\multirow{2}{*}{GMAC/s} & \multirow{2}{*}{CSIG$\uparrow$}   &\multirow{2}{*}{CBAK$\uparrow$}     &\multirow{2}{*}{COVL$\uparrow$}   & \multirow{2}{*}{PESQ$\uparrow$}    & \multirow{2}{*}{STOI$\uparrow$}     & \multicolumn{2}{c}{CER (\%)$\downarrow$}         \\ 
    & &(M) & & & & & & & DS2 & W2V2 \\ \midrule
    A1 &ROSE & 36.98 & 4.73 & 4.58$\pm$.04 & 3.22$\pm$.04 & 3.85$\pm$.03 & 3.09$\pm$.03 & 0.84$\pm$.00 & 3.50$\pm$.18 & 2.42$\pm$.19\\
    \cmidrule{1-11}
    B1 & A1 $w/o$ $L_{MAE}$    & 36.98 & 4.73 & 4.32$\pm$.03   & 2.99$\pm$.04  & 3.51$\pm$.03  & 2.73$\pm$.04  & 0.82$\pm$.00 & 3.88$\pm$.20  & 2.45$\pm$.19        \\
    B2 & A1 $w/o$ $L_{MAG}$    & 36.98 & 4.73 & 4.36$\pm$.03   & 3.03$\pm$.04  & 3.54$\pm$.04  & 2.75$\pm$.04  & 0.82$\pm$.00 & 3.86$\pm$.20  & 2.47$\pm$.20        \\
    B3 & A1 $w/o$ $L_{SE}$    & 36.98 & 4.73 & 4.28$\pm$.04   & 2.96$\pm$.04  & 3.48$\pm$.04  & 2.69$\pm$.04  & 0.81$\pm$.00 & 3.91$\pm$.22   & 2.50$\pm$.19        \\
    B4 & A1 $w/o$ $L_{Spec.}$  & 36.98 & 4.73 & 4.42$\pm$.03   & 3.08$\pm$.04  & 3.64$\pm$.04  & 2.85$\pm$.04  & 0.83$\pm$.00 & 4.00$\pm$.19  & 2.53$\pm$.20          \\
    B5 & A1 $w/o$ $L_{MFCC}$  & 36.98 & 4.73 & 4.40$\pm$.04   & 3.04$\pm$.04  & 3.64$\pm$.04  & 2.88$\pm$.04  & 0.83$\pm$.00 & 4.08$\pm$.21   & 2.56$\pm$.23       \\
    B6 & A1 $w/o$ $L_{ASR}$   & 36.98 & 4.73 & 4.33$\pm$.04   & 3.01$\pm$.03  & 3.53$\pm$.04  & 2.71$\pm$.04  & 0.82$\pm$.00  & 4.29$\pm$.24  & 2.60$\pm$.24       \\ 
    \cmidrule{1-11}
    C1 & B6 $w/o$ ABSF        & 35.80 & 4.18 & 4.27$\pm$.04   & 2.94$\pm$.04  & 3.46$\pm$.04  & 2.66$\pm$.04  & 0.82$\pm$.00  & 4.31$\pm$.24  & 2.63$\pm$.24       \\ 
    C2 & B6 $w/o$ CSAtt       & 35.40 & 4.72 & 4.15$\pm$.03   & 2.84$\pm$.04  & 3.33$\pm$.04  & 2.51$\pm$.04  & 0.80$\pm$.00  & 4.36$\pm$.24  & 2.68$\pm$.28        \\ 
    C3 & B6 $w/o$ $attn.$     & 34.22 & 4.17 & 4.04$\pm$.04   & 2.76$\pm$.04  & 3.21$\pm$.04  & 2.39$\pm$.03  & 0.79$\pm$.00  & 4.40$\pm$.26  & 2.71$\pm$.27        \\
    \bottomrule
    \end{tabular}
\end{table*}

\begin{table*}[!t]
    \caption{SE results on Voice Bank + DEMAND dataset. \label{tab:VCTKresults}“—" denotes the results are not provided in the associated papers. “*" denotes that the results are reproduced by us. Optimal values are \textbf{bold} and sub-optimal values are marked with \uwave{underlined wavy line} in each column.}
    \centering
    \begin{tabular}{lccc|ccccc}
    \toprule
    Model   & Input  & \#Param. (M) & GMAC/s  &CSIG$\uparrow$  & CBAK$\uparrow$  & COVL$\uparrow$  & PESQ$\uparrow$  & STOI$\uparrow$   \\ \midrule
    Noisy                               & -             &-  &- & 3.35  & 2.44  & 2.63  & 1.97  & 0.92    \\
    \cmidrule{1-9}
    \multicolumn{9}{c}{Traditional Methods}  \\
    \cmidrule{1-9}
    Wiener \cite{RN162}                 & Waveform      &-  &-  & 3.23  & 2.68  & 2.67  & 2.22  & 0.93    \\
    Spec-Sub \cite{RN102} *             & Magnitude     &-  &-  & 3.40  & 2.71  & 2.74  & 2.26  & 0.93    \\
    \cmidrule{1-9}
    \multicolumn{9}{c}{GAN-based Generative Methods}  \\ 
    \cmidrule{1-9}
    SEGAN \cite{RN100}                  & Waveform      &65.82  &13.39  & 3.48  & 2.94  & 2.80  & 2.16  & —     \\
    MetricGAN \cite{fu2019metricgan}    & Magnitude     &1.86  &28.51  & 3.99  & 3.18  & 3.42  & 2.86  & —     \\
    CRGAN  \cite{crgan}                 & Magnitude     &—  &—  & 4.16  & 3.24  & 3.54  & 2.92  & 0.94    \\
    MetricGAN+ \cite{meganplus}         & Magnitude     &1.86  &28.51  & 4.14  & 3.16  & 3.64  & 3.15 & —         \\
    MetricGAN-OKD (PE+CS) \cite{shin2023metricgan}       
                                        & Magnitude     &0.82  &—  & 4.23  & 3.07  & 3.73  & \textbf{3.24}  & —     \\
    \cmidrule{1-9}
    \multicolumn{9}{c}{Diffusion-based Generative Methods}  \\ 
    \cmidrule{1-9}
    DiffuSE \cite{diffuse} *             & Magnitude     &2.23  &210.31  & 3.64  & 2.85  & 3.02  & 2.43  & 0.90    \\
    CDiffuSE \cite{cdiffuse} *           & Magnitude     &4.28  &190.23  & 3.85  & 3.03  & 3.20  & 2.45  & 0.91    \\
    SGMSE \cite{sgmse} *                 & Magnitude     &3.89  &152.77  & 4.30  & 3.52  & 3.67  & 2.93  & 0.94    \\
    \cmidrule{1-9}
    \multicolumn{9}{c}{Discriminative Methods}  \\ 
    \cmidrule{1-9}
    CRN \cite{crn}                      & Magnitude     &1.70  &8.80  & 3.59  & 3.11  & 2.71  & 2.48  & 0.93    \\ 
    DEMUCS \cite{RN148}                 & Waveform      &68.87  &38.12  & 4.31  & 3.40  & 3.63  & 3.07  & 0.95    \\
    DCCRN-E \cite{dccrn}                & RI            &3.67  &33.48  & 3.74  & 3.13  & 2.75  & 2.54  & 0.94    \\
    PHASEN \cite{yin2020phasen}         & Phase+Magnitude 
                                                        &—  &—  & 4.21  & 3.55  & 3.62  & 2.99 & —      \\

    TSTNN \cite{wang2021tstnn}          & Waveform      &0.92  &—  & 4.17  & 3.53  & 3.49  & 2.96  & 0.95    \\ 
    SE-Conformer \cite{kim2021se}       & Waveform      &75.28  &44.96  & \uwave{4.45}  & 3.55  & \textbf{3.82} & 3.13  & 0.95    \\ 
    DB-AIAT \cite{yu2022dual} *         & RI+Magnitude  &2.81  &41.8  & 4.41  & 3.53  & 3.76  & \uwave{3.21}  & 0.95    \\
    FullSubNet+ \cite{fullsubnet}       & RI+Magnitude  &8.67  &48.57  & 3.86  & 3.42  & 3.57  & 2.88  & 0.94    \\
    GaGNet \cite{li2022glance}          & RI+Magnitude  &—  &—  & 4.26  & 3.45  & 3.59  & 2.94  & 0.95  \\ 
    DeepFilterNet2 \cite{dfn2}          & RI+Magnitude  &2.31  &—  & 4.30  & 3.40  & 3.70  & 3.08  & 0.94    \\
    MANNER-S-8.1GF \cite{shin2022multi} & Waveform      &1.38  &4.09  & 4.40  & 3.55  & 3.74  & 3.01  & 0.95     \\
    BSRNN \cite{Yu2023HighFS} *         & Magnitude     &2.96  &2.57  & 4.41  & \textbf{3.59}  & \uwave{3.80}  & 3.05  & 0.95    \\
    TF-GridNet \cite{wang2023tf} *      & RI            &8.2  &19.2  & 4.37  & 3.50  & 3.74  & 2.98  & 0.94    \\
    \cmidrule{1-9}
    ROSE (Ours)                         & Waveform      &36.98  &4.73  & \textbf{4.47} & \uwave{3.56} & 3.72  & 3.01  & \textbf{0.95} \\ 
    \bottomrule
\end{tabular}
\end{table*}

\section{Results and Discussions}
\subsection{Performance Evaluation}
The objective and subjective experimental results for the performance comparison are shown in Table \ref{tab:ATCresults_1} and Table \ref{tab:ATCresults_2}, respectively. The following conclusions can be obtained from the experimental results:

1) The traditional SE methods (Wiener and Spec-Sub) fail to suppress the speech echo, and even cause extra noise to deteriorate the quality of the original speeches. The results also indicate that the speech echo is a dedicated phenomenon in the ATC domain, which cannot be addressed by the traditional denoising method.

2) Although the GAN-based and Diffusion-based generative methods improve the objective and subjective SE-related metrics of the noisy speech, they deteriorate the ASR accuracy due to the generation mechanism for pseudo-speech by learning the data distribution of clean signals. The main purpose is to satisfy the auditory requirements of the speech quality but fails to enclose ASR-required features to achieve desired recognition accuracy.

3) According to the discriminative models, although the time-domain methods are slightly inferior to the T-F methods in terms of SE-related metrics, they have a significant advantage in ASR performance. The results show that the original waveform contains the complete information of the speech, while the input features (e.g., magnitude, phase, complex and their combinations) of the T-F methods may fail to present the same complete features as the waveform, further leading to the lack of critical ASR-related linguistic information, especially in ATC scenarios.

4) Compared to the causal models, the non-causal models obtain superior performance on both SE and ASR tasks. In general, the non-causal paradigm can leverage the sequential information in the past and future dimensions to capture the long-distance temporal dependencies among the signal frames.

5) By focusing on the ASR performance, almost all the comparative baselines suffer inferior CER compared to the noisy speeches (only marginal improvements in certain models) since they are SE-oriented approaches to improve the auditory quality. The metrics of AIR indicate that improving the clarity and intelligibility of speech can indeed improve the rater's ability to understand ATC instructions.

In summary, the proposed approach aims to address the ATC speech echo and achieves the desired performance improvement on the real-world dataset. Although the proposed ROSE only achieves sub-optimal computational costs, fortunately, it generally achieves the best performance in ATC scenarios, which is the preferred goal of this work. Furthermore, the proposed model achieves improved recognition performance both for the feature-adapted DS2 model and the SSL-based W2V2 model, which further confirms the generalization of the proposed technical improvements.

\subsection{Ablation Study}
To validate the effectiveness of the multi-objective learning mechanism, ablation experiments are conducted by combining different loss functions in the proposed model. The results are reported in Table \ref{tab:ATCresults_ablation} (group B). Benefiting from the objective-specific optimization, the SE-oriented losses guide the model to achieve better SE-related performance than that of the ASR-oriented losses, while the two methods (B6 vs. B3) present different trends in the recognition performance. In B3, the SE-related metrics show a large performance degradation due to the removal of the SE-oriented losses, and the ASR performance also suffers from the higher CER (A1 vs. B3). Compared with experiments B1$\sim$B3, the spectrogram loss can contribute to the desired performance improvement for the ASR task by optimizing the spectral distance between the enhanced signal and the reference signal, further enhancing the effective acoustic features and suppressing the interference features (B4). Compared to the spectrogram features, the MFCC-based SC loss achieves better improvement since it is a more sophisticated handcrafted feature (B4 vs. B5). The best performance is reached by combining the two SC losses, in which DS2-CER and W2V2-CER are improved from 4.29\% to 3.50\% and from 2.60\% to 2.42\% (A1 vs. B6), respectively.

By incorporating the SC loss into ASR-required features learning, the ASR performance can be considerably enhanced by the multi-objective framework without any model retraining, which validates the core motivation of this work. Except for the SE performance, the proposed model also harvests higher ASR performance, which can be attributed that the multi-objective loss learns the common representations across the two optimization objectives thanks to its powerful ability on feature learning.

To validate the ability of the proposed attention modules, ablation experiments are conducted by combining different attention modules in the single-objective model (B6). As can be seen from Table \ref{tab:ATCresults_ablation} (group C), both ABSF and CSAtt modules can enhance the model performance in terms of both the SE- and ASR-related metrics. Specifically, compared to the ABSF module, the CSAtt can achieve higher performance improvement by optimizing the learned features, where the first four SE-related metrics are improved by about 0.13 on average, and both DS2-CER and W2V2-CER are decreased by 0.05\% (C1 vs. C2). To be specific, the attention module focuses on informative feature channels to suppress noise interference, while highlighting selective frames to fit the data distribution between features and text labels. Finally, the proposed model can achieve the best performance by combining the two attention modules (B6 vs. C3), which validates the proposed model architecture.

\begin{table}[!t]
    \caption{ASR results on two artificial datasets. “$a$” denotes additive noise, “$e$” denotes simulative echo. \label{tab:asr}}
    \centering
    \begin{tabular}{l|cccc}
    \toprule
    \multirow{2}{*}{Model}       & \multicolumn{2}{c}{DS2} & \multicolumn{2}{c}{W2V2}\\
    & $a$CER(\%)$\downarrow$   & $e$CER(\%)$\downarrow$   & $a$CER(\%)$\downarrow$  & $e$CER(\%)$\downarrow$  \\ 
    \midrule
    Mixed                                & 33.36   & 42.72 & 7.03    & 10.22 \\
    SE-Conformer                         & 21.43   & 31.10 & 6.54    & 8.12  \\
    DCCRN-E                              & 26.69   & 33.67 & 6.88    & 9.72  \\
    BSRNN                                & 23.79   & 32.23 & 6.73    & 8.55  \\
    \cmidrule{1-5}
    ROSE (Ours)                          & \textbf{12.75}   & \textbf{22.31}   & \textbf{4.39}    & \textbf{6.55} \\ 
    \quad $w/o$ $L_{ASR}$                & 19.12   & 29.44  & 6.29  & 7.99 \\
    \quad  $w/o$ $attn.$                 & 17.86   & 26.56  & 5.57  & 7.22 \\ 
    
    \cmidrule{1-5}
    test-clean                           & \multicolumn{2}{c}{9.97} & \multicolumn{2}{c}{2.49}\\
    \bottomrule
    \end{tabular}
  \end{table}

\begin{figure*}[!t]
  \centering
  \begin{minipage}[t]{0.9\linewidth}
  \centering
      \begin{tabular}{@{\extracolsep{\fill}}c@{}c@{}c@{}@{\extracolsep{\fill}}}
          \includegraphics[width=0.33\linewidth]{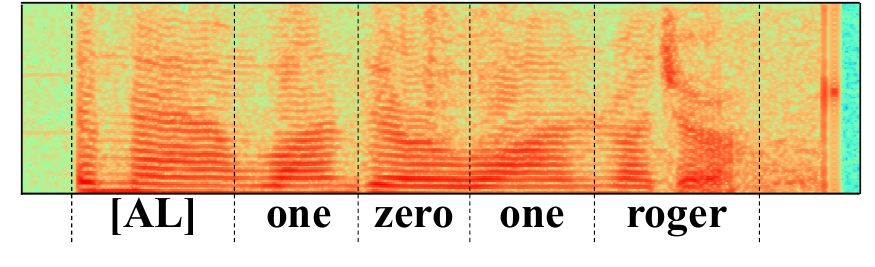} &
          \includegraphics[width=0.33\linewidth]{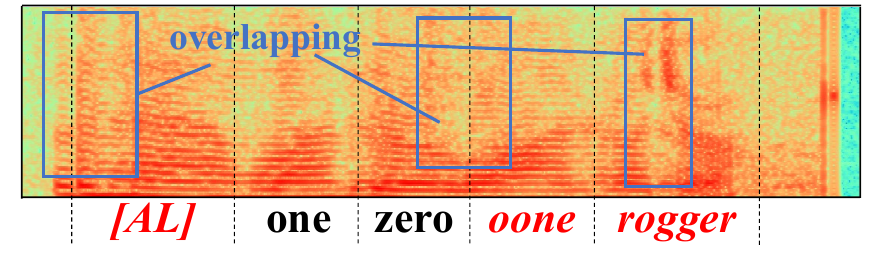}&
          \includegraphics[width=0.33\linewidth]{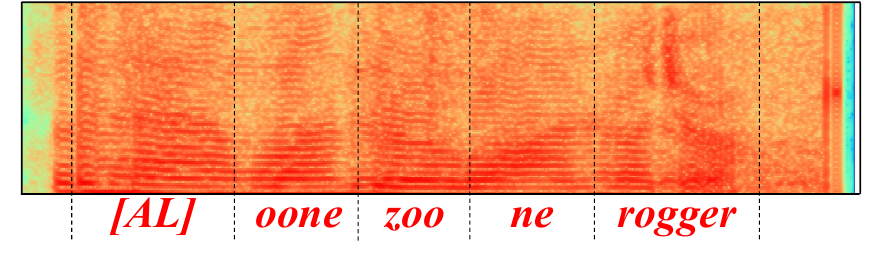}\\
          (a) Clean & (b) Noisy & (c) Wiener\\
      \end{tabular}
  \end{minipage}
  \begin{minipage}[t]{0.9\linewidth}
  \centering
      \begin{tabular}{@{\extracolsep{\fill}}c@{}c@{}c@{}@{\extracolsep{\fill}}}
          \includegraphics[width=0.33\linewidth]{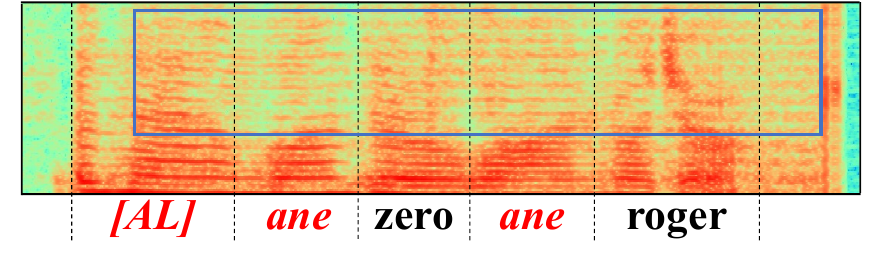} &
          \includegraphics[width=0.33\linewidth]{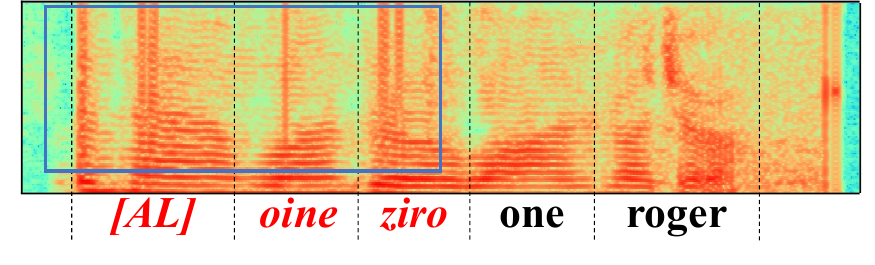} &
          \includegraphics[width=0.33\linewidth]{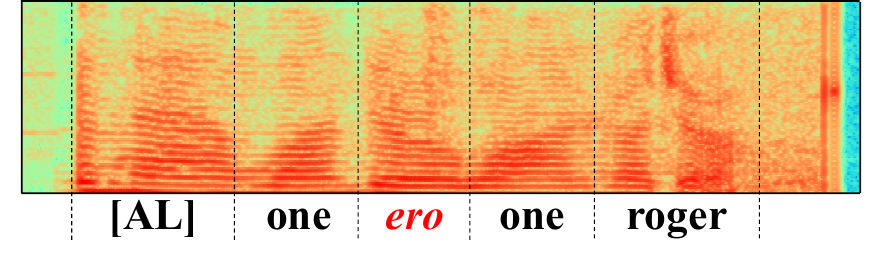}\\
          (d) CRN & (e) DCCRN-E & (f) FullSubNet+\\
      \end{tabular}
  \end{minipage}
  \begin{minipage}[t]{0.9\linewidth}
  \centering
      \begin{tabular}{@{\extracolsep{\fill}}c@{}c@{}c@{}@{\extracolsep{\fill}}}
          \includegraphics[width=0.33\linewidth]{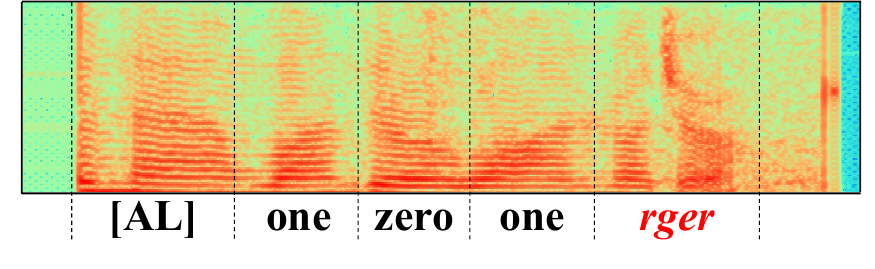} &
          \includegraphics[width=0.33\linewidth]{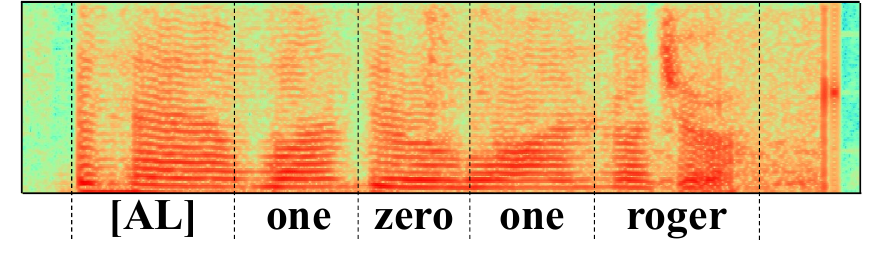}&
          \includegraphics[width=0.33\linewidth]{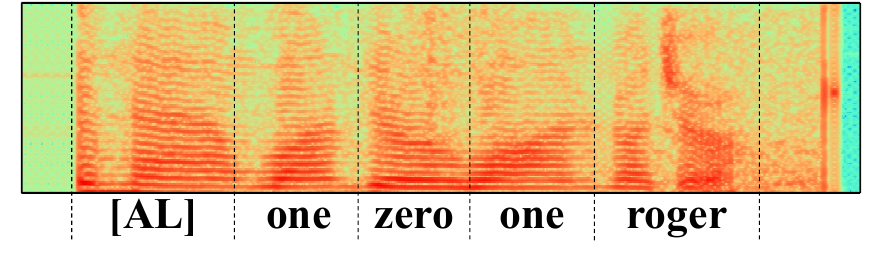}\\
          (g) DEMUCS & (h) SE-Conformer & (i) ROSE (Ours)\\
      \end{tabular}
  \end{minipage}
  \caption{Visualization for the Spectrograms and ASR results.}
  \label{fig:spectrogram_asr}
\end{figure*}

\subsection{Experiments on public datasets}
\subsubsection{Experiments for the SE}
To validate the SE performance of the proposed model on the public dataset, in this section, several representative SOTA models with different inputs are selected to compare with ROSE on Voice Bank + DEMAND dataset \cite{valentini2016investigating}. Specifically, five types of input are classfied as: \emph{1) Waveform},  \emph{2) Magnitude}, \emph{3) Phase + Magnitude}, \emph{4) Real-imaginary (RI)} and \emph{5) RI + Magnitude}.

In the Voice Bank + DEMAND dataset, the clean speeches were recorded from 30 speakers, 28 of which are used for the training set and 2 for the test set. To formulate the noisy speeches in the training set, a total of 10 noise models (2 artificial and 8 from the DEMAND dataset \cite{RN215}) with 4 signal-to-noise-ratio (SNR) levels (0, 5, 10, and 15 dB) are performed on clean speeches (28 speakers) to generate artificial data pairs to support the SE task. To generate an independent test set, 5 unseen noise models (all from the DEMAND dataset) with 4 SNR levels (2.5, 7.5, 12.5, and 17.5 dB) are performed to clean speeches (2 speakers) to obtain the noisy test set. Finally, the training set and test set are 11572 and 824 speech pairs respectively. All samples are clipped to 4 seconds long and audio samples are made available online\footnote[3]{\url{https://github.com/XCYu-0903/ROSE}}.

For the experimental configurations, the setups of the ROSE are the same as mentioned in Section \ref{sc:Configurations}. The experimental results are reported in Table \ref{tab:VCTKresults}. Although ROSE does not achieve the best performance on COVL and PESQ, it still achieves optimal performance on CSIG and STOI and sub-optimal performance on CBAK. Inside the MetricGAN-related methods, due to the PESQ-oriented optimization target involved in the training procedure, the models have achieved cliff-like superiorities on PESQ, while other metrics are inferior. This result shows that the objective-oriented loss function can improve task-oriented feature learning. In addition, it can also be found that the traditional methods (Wiener and Spec-Sub) can achieve better performance on the public dataset with artificial noise models, which also supports the research specificities of the ATC speech corpus with real-world noise and speech echo phenomenon.

\subsubsection{Experiments for the ASR}
To validate the ASR performance of ROSE on the public dataset, according to the mixed type of signals, two datasets based on test-clean samples in the Librispeech dataset \cite{panayotov2015librispeech} are artificially synthesized, which are additive noise and simulative echo datasets. Specifically, the mixed noisy samples of the additive noise dataset are formed by mixing each utterance of test-clean samples with the noises from the test set of DNS benchmark \cite{reddy2020interspeech} with 4 SNR levels (-3, 0, 3, and 6 dB). 

To simulate as much as possible the speech echo in the ATC domain, each utterance of test-clean samples is copied twice to reproduce the ATC communication process. The samples are mixed with 30dB and 10dB Gaussian white noises to simulate the sent and more complex received speeches (mentioned in Figure \ref{fig:background}) respectively. Finally, the mixed noisy samples of the simulative echo dataset are formed by overlapping the above two samples with a random temporal delay between 10 and 200 milliseconds and clipped to ensure the temporal alignment with test-clean samples. Note that the test-clean samples are treated as echo-free speeches for comparison with the additive noise dataset.

Comparative models are trained on Voice Bank + DEMAND dataset to enhance blind artificial datasets (i.e., SE-Conformer, DCCRN-E, BSRNN and ROSE). Two ASR models are applied to evaluate the enhanced artificial set, including a supervised DS2 model and an SSL-based W2V2 model. Specifically, the DS2 model is trained on Librispeech 960-hours training dataset, in which the input feature is the 81-dimension spectrogram. The pre-trained W2V2 model provided on fairseq\footnote[4]{The BASE W2V2 model pre-trained on Librispeech 960-hours training set is used.} is fine-tuned on Librispeech dev-clean set with the CTC loss function. All the above ASR models are end-to-end models without language models.

The experimental results are reported in Table \ref{tab:asr}. Speech is a continuous signal with strong temporal correlations between adjacent frames. Compared to additive noise, signal overlapping severely degrades the discriminability of text-related features, leading to collapsed representations \cite{wdtcn,evae}. Similar to results in Table \ref{tab:ATCresults_1}, compared to the T-F domain approaches, the time-domain approaches achieve better performance in additive noise and simulated echo scenarios. The proposed attention modules and multi-objective learning mechanism achieve performance improvements on all datasets, and multi-objective learning can provide higher CER reduction, confirming the generalizability of the proposed technical improvements on different kinds of speech noise.

\subsection{Visualization}
\subsubsection{Enhancement Results}
To better demonstrate the performance of each model on SE and ASR tasks, we select a pure English example from the test set of the ATC corpus (the selected speech text ground-truth is: \emph{\textbf{[AL] one zero one roger}}, in which [AL] refers to the airline) and draw the spectrums in different cases.

As shown in Figure \ref{fig:spectrogram_asr}, compared with the clean spectrogram (a), the noisy spectrogram (b) has obvious overlapping phenomena, resulting in blurred gaps between English word pronunciations. Wiener (c) not only fails to effectively remove noise components but also brings redundant signals to speech, resulting in inferior ASR performance.

There are some defects in the selected T-F domain methods. The high-frequency area of CRN (d) shows sparseness leading to collapsed features. The sequence of DCCRN-E (e) has unexpected mutational signals, which further affect the stationarity of the speech. On the contrary, the time domain methods achieve superior performance, especially the ROSE (i) obtains the result closest to clean (a).

\begin{figure*}[!t]
    \centering
    \begin{minipage}[t]{0.9\linewidth}
    \centering
        \begin{tabular}{@{\extracolsep{\fill}}c@{}c@{}c@{}@{\extracolsep{\fill}}}
            \includegraphics[width=0.33\linewidth]{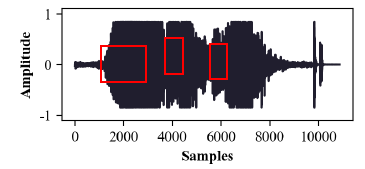}&
            \includegraphics[width=0.33\linewidth]{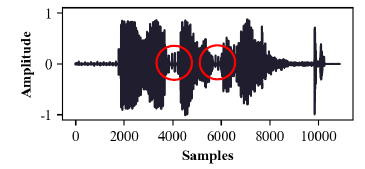}&
            \includegraphics[width=0.33\linewidth]{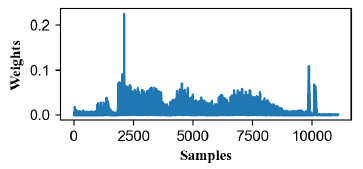}\\
            (a) Noisy & (b) Enhanced (without ABSF) & (c) Attention mask of the first layer \\
        \end{tabular}
    \end{minipage}
    \begin{minipage}[t]{0.9\linewidth}
    \centering
        \begin{tabular}{@{\extracolsep{\fill}}c@{}c@{}c@{}@{\extracolsep{\fill}}}
            \includegraphics[width=0.33\linewidth]{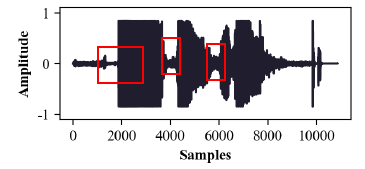}&
            \includegraphics[width=0.33\linewidth]{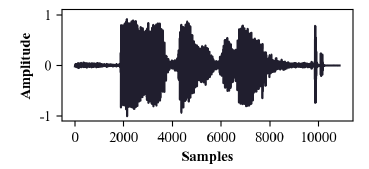}&
            \includegraphics[width=0.33\linewidth]{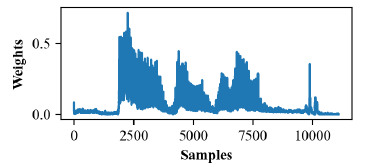}\\
            (d) Clean & (e) Enhanced (with ABSF) & (f) Attention mask of the last layer \\
        \end{tabular}
    \end{minipage}
    \caption{Visualization for the ABSF module.}
    \label{fig:absf}
\end{figure*}

\begin{figure*}[!t]
    \centering
    \begin{minipage}[t]{0.9\linewidth}
    \centering
        \begin{tabular}{@{\extracolsep{\fill}}c@{}c@{}c@{}@{\extracolsep{\fill}}}
            \includegraphics[width=0.33\linewidth]{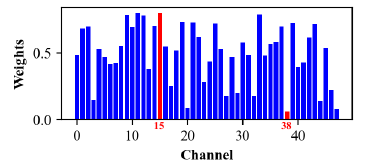}&
            \includegraphics[width=0.33\linewidth]{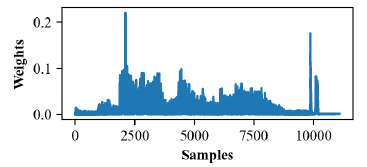}&
            \includegraphics[width=0.33\linewidth]{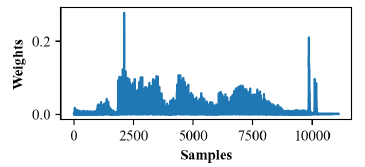}\\
            (a) Channel weights of the first layer & (b) Attn. maps of the 38$^{th}$ channel & (c) Attn. maps of the 15$^{th}$ channel\\
        \end{tabular}
    \end{minipage}
    \begin{minipage}[t]{0.9\linewidth}
        \centering
            \begin{tabular}{@{\extracolsep{\fill}}c@{}c@{}c@{}@{\extracolsep{\fill}}}
                \includegraphics[width=0.33\linewidth]{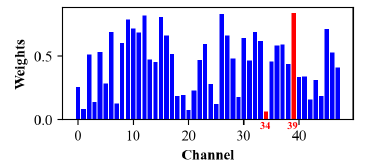}&
                \includegraphics[width=0.33\linewidth]{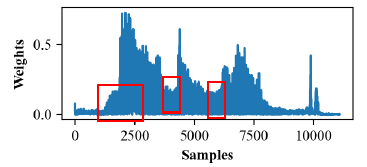}&
                \includegraphics[width=0.33\linewidth]{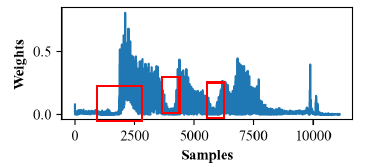}\\
                (d) Channel weights of the last layer& (e) Attn. maps of the 34$^{th}$ channel& (f) Attn. maps of the 39$^{th}$ channel\\
            \end{tabular}
        \end{minipage}
        \caption{Visualization for the channel attention of the CSAtt module.}
    \label{fig:csatt_channel}
\end{figure*}

\begin{figure*}[!t]
    \centering
    \begin{minipage}[t]{0.9\linewidth}
        \centering
            \begin{tabular}{@{\extracolsep{\fill}}c@{}c@{}c@{}@{\extracolsep{\fill}}}
                \includegraphics[width=0.33\linewidth]{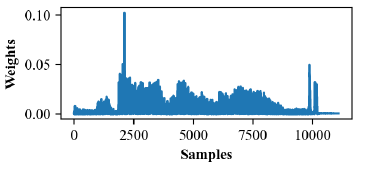}&
                \includegraphics[width=0.33\linewidth]{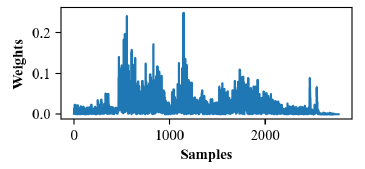}&
                \includegraphics[width=0.33\linewidth]{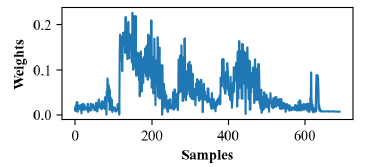}\\
                (a) 1$^{st}$ layer of the Encoder& (b) 3$^{rd}$ layer of the Encoder & (c) 5$^{th}$ layer of the Encoder\\
            \end{tabular}
        \end{minipage}
    \begin{minipage}[t]{0.9\linewidth}
        \centering
            \begin{tabular}{@{\extracolsep{\fill}}c@{}c@{}c@{}@{\extracolsep{\fill}}}
                \includegraphics[width=0.33\linewidth]{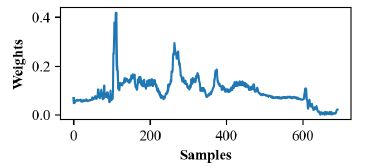}&
                \includegraphics[width=0.33\linewidth]{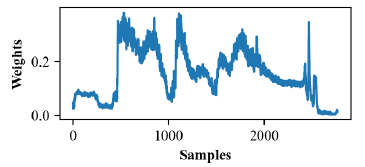}&
                \includegraphics[width=0.33\linewidth]{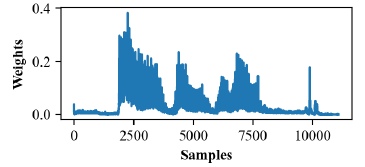}\\
                (d) 5$^{th}$ layer of the Decoder& (e) 3$^{rd}$ layer of the Decoder& (f) 1$^{st}$ layer of the  Decoder\\
            \end{tabular}
        \end{minipage}
        \caption{Visualization for the sequence attention of the CSAtt module.}
    \label{fig:csatt_sequence}
\end{figure*}

\subsubsection{The ABSF Module}
\label{sc:absf_visual}
To validate the performance of the ABSF module, we select an example from the test set and visualize it in the following cases: clean waveform, noisy waveform, enhanced waveform (with ABSF or not), and the attention mask $A_{i}$ (mentioned in Section \ref{sc:absf}) of the first layer of the encoder (shallowest) and the last layer of the decoder (deepest). All feature representatives in this work are adaptively adjusted for visualization purposes.

As shown in Figure \ref{fig:absf}, the final attention mask (f) behaves as learning a voice activity detector that preserves features as much as possible to focus on active regions of the voice. Compared with the result (b) obtained by directly concatenating the hierarchical features in skip connections, the ABSF module (e) improves the feature space of the model, obtaining the enhanced signal that is closer to the clean signal (d). According to (d), (e) and (f), we can find that the learned attention mask is similar to the positive amplitude profile of the clean waveform, achieving that by assigning higher weights to valid voice activity frames and lower weights to redundant frames, further validating the fusion processing between encoders and decoders.

\subsubsection{Channel Attention of the CSAtt Module}
Since spoken words and noise are distributed in different speech feature channels, the purpose of the channel attention of the CSAtt module is to give higher weights to the channels with rich spoken information and lower weights to those with noise.

To confirm the channel attention, we visualize the channel attention weights $W_{C}$ (mentioned in Section \ref{sc:csatt}) obtained from the channel path of the CSAtt module of the first layer of the encoder and the last layer of the decoder, respectively. We further visualize the attention maps $X_{C}$ corresponding to the channel dimensions in $W_{C}$ that are assigned the highest and lowest weights. The visualization sample is the same as that of in ABSF module (as shown in Figure \ref{fig:absf}).

As shown in Figure \ref{fig:csatt_channel}, (a) and (d) represent the distribution of channel attention weights in the selected network layers (hidden dimensions is $H=48$), respectively, in which assigned highest and lowest weights of channel indexes are marked in red. (b) and (c) are the attention maps of the assigned lowest weight (indexed as 38) and highest weight (indexed as 15) in (a), respectively. while (e) and (f) are the attention maps of the assigned lowest weight (indexed as 15) and highest weight (indexed as 20) in (d), respectively. In general, in the first encoder layer, (b) and (c) present a similar distribution in the different attention channels, which confuses the model to learn effective spoken information in the shallow layer of the neural network. On the contrary, in the deeper decoder layer, (e) and (f) learn the discriminative attention weights to suppress the noisy features marked by red rectangles, in which the corresponding time-domain samples are also marked in Figure \ref{fig:absf} (a) and (d). The visualization also confirms that channel attention achieves the assignment of the expected weight for the SE task.

\subsubsection{Sequence Attention of the CSAtt Module}
The sequence attention of the CSAtt module is required to focus on speech central frames with valid phonemes while filtering marginal echo frames. In the feature domain, the process of sequence attention weighting the frames is the process of continuously fitting the feature maps to the clean distribution.

To intuitively understand the learning process of the sequence attention mechanism, we visualize the sequence attention maps $X_{L}$ (mentioned in Section \ref{sc:csatt}) obtained from the sequence path of the CSAtt module of the first, third, and fifth encoder layers and their corresponding decoder layers, where the input utterance is the same as that in ABSF visualization.

As shown in Figure \ref{fig:csatt_sequence} (layers from shallow to deep), in the shallow layer (a), we observe that attention maps present a mixture of clean and noisy features, which fail to distinguish informative signals. In (b), the result roughly marks the area where valid information may appear, but it is still not clear. The weights assignment of valid signals is gradually revised from layer (c) to layer (d) until layer (e) achieves convergence, which shows that the ability to reconstruct the clean features is improved with the stacking of sequence attention.

\begin{table}[!t]
    \caption{Experimental results for AEC and dereverberation tasks. \label{tab:AEC_de}}
    \centering
    \begin{tabular}{l|ccc|c}
    \toprule
    Model           & PESQ$\uparrow$ &CD$\downarrow$ &SRMR$\uparrow$ & DS2-CER$\downarrow$ \\ \midrule
    Noisy           & 1.80$\pm$.01  & 3.66$\pm$.06  & 2.24$\pm$.02  & 4.66$\pm$.24  \\
    \cmidrule{1-5}
    NLMS \cite{RN210}                       & 1.50$\pm$.03  & 4.82$\pm$.09  & 1.55$\pm$.04  & 12.89$\pm$.46  \\
    Align-CRUSE \cite{indenbom2022deep}     & 1.84$\pm$.03  & 4.07$\pm$.10  & 1.68$\pm$.03  & 10.99$\pm$.38  \\
    NKF  \cite{yang2023low}                 & 1.93$\pm$.03  & 3.89$\pm$.09  & 1.93$\pm$.04  & 10.52$\pm$.36  \\
    \cmidrule{1-5}
    WPE \cite{yoshioka2012generalization}   & 1.55$\pm$.04  & 4.77$\pm$.10  & 1.61$\pm$.04  & 11.67$\pm$.39 \\
    WD-TCN \cite{wdtcn}                     & 2.78$\pm$.04  & 2.65$\pm$.03  & 4.04$\pm$.04  & 6.09$\pm$.27  \\
    RVAE-EM-S \cite{evae}                   & 2.81$\pm$.04  & 2.61$\pm$.02  & 4.23$\pm$.03  & 5.11$\pm$.25  \\
    \cmidrule{1-5} 
    ROSE (Ours)    & \textbf{3.09$\pm$.03}    & \textbf{2.12$\pm$.05}	& \textbf{5.33$\pm$.19} & \textbf{3.50$\pm$.18}  \\ 
    \bottomrule
    \end{tabular}
\end{table}

\begin{table}[!t]
    \caption{Experimental results of integrating the proposed ABSF, CSAtt and $L_{ASR}$ into SE-Conformer. \label{tab:se-conformer}}
    \centering
    \begin{tabular}{ccc|cc}
    \toprule
    ABSF           & CSAtt       & $L_{ASR}$    &PESQ$\uparrow$  & DS2-CER$\downarrow$ \\ \midrule
    \usym{2613}    &\usym{2613}  & \usym{2613}   & 3.02$\pm$.04  & 4.28$\pm$.20  \\
    \usym{2613}    &\usym{2613}  & \Checkmark    & 3.05$\pm$.04  & 4.19$\pm$.19  \\
    \Checkmark     &\usym{2613}  & \usym{2613}   & 3.06$\pm$.03  & 4.24$\pm$.21  \\
    \Checkmark     &1   & \usym{2613}    & 3.07$\pm$.04  & 4.19$\pm$.20  \\
    \Checkmark     &3   & \usym{2613}    & 3.09$\pm$.04  & 4.15$\pm$.19  \\
    \Checkmark     &5   & \usym{2613}    & 3.04$\pm$.05  & 4.21$\pm$.19  \\
    \Checkmark     &3   & \Checkmark     & 3.11$\pm$.04  & 4.05$\pm$.19  \\
    \bottomrule
    \end{tabular}
\end{table}

\subsection{Discussions}
\subsubsection{Confirmation of the specific natures of the speech echo in the ATC domain}
As mentioned in Section \ref{sc:introduction}, the type of signal that causes the speech echo cannot be defined with certainty. Therefore, we additionally use acoustic echo cancellation (AEC) methods and dereverberation methods to explore the effectiveness of different signal processing methods in eliminating the speech echo. The clean-noisy pair dataset is used to conduct relevant experiments.

For the AEC methods, three open-source baselines are selected to conduct the experiments, including a traditional method NLMS \cite{RN210} and two DNN-based methods Align-CRUSE \cite{indenbom2022deep} and NKF \cite{yang2023low}. According to previous work \cite{WOS:001058793500002}, we can only collect the received speech (clean) and the combined (noisy) speech in a real-world ATC environment. To simulate the far-end signal and not introduce other signals, we generate the simulated sent speech only by applying random time offsets between 0 and 200 milliseconds to the received speech and then clip it to ensure the temporal alignment with the original signal. Therefore, to fit the AEC scenarios, we regard the received speech as the near-end signal, the simulated sent speech as the far-end reference signal, and the combined speech (with the speech echo) as the mixed signal. The purpose of the AEC task is to remove the far-end signal from the mixed signal to obtain the estimated near-end signal, which is also the same as the purpose of the SE task. For dereverberation methods, three open-source baselines are selected to conduct the experiments, including a traditional method WPE \cite{yoshioka2012generalization} and two DNN-based methods WD-TCN \cite{wdtcn} and RVAE-EM-S \cite{evae}. To fit the dereverberation scenarios, we regard the received speech as the clean signal and combined speech as the reverberant signal.

For the metrics of the above tasks, we adopt the objective evaluation metrics on the test set, including PESQ \cite{RN212}, cepstral distance (CD), speech-to-reverberation modulation energy ratio (SRMR) \cite{kothapally2020skipconvnet} and DS2-CER to measure effects of the signals. Except for SRMR, the calculation of other metrics requires the corresponding reference speech. For CD and DS2-CER, lower values indicate better performance; for other metrics, higher values indicate better performance. We also report the mean results and their 95\% confidence interval.

As shown in Table \ref{tab:AEC_de}, the AEC-related methods not only fail to remove the speech echo, but also lead to worse speech quality. The reason is that in the common AEC scenarios, the far-end and near-end signals are different signals, and AEC methods are able to achieve signal separation based on the reference signals. In the ATC scenarios, the far-end and near-end signals are highly similar (almost the same, and all signals are issued by the same ATCO), and the AEC methods are unable to capture effective features to support SE and ASR tasks. Similar to the traditional methods of SE, the WPE model fails to remove the speech echo and even obtains worse performance than noisy speech. Although the DNN-based dereverberation methods can obtain the desired SE performance due to the task objective with the SE, they still suffer from poor ASR performance even compared to the noisy speech, which fails to meet the requirements for both the SE and ASR tasks in this work. Therefore, the experimental results confirm that the speech echo is a specific mechanism that cannot be removed by common AEC and dereverberation methods, which also supports the motivation of our proposed ROSE framework.

\subsubsection{Exploration for the Generalizability of the Proposed Technical Improvements}
Previous experimental results show that all the proposed technical improvements (i.e., ABSF module, CSAtt module and ASR-oriented loss function) achieve the desired performance on the architecture used in this work. As a similar U-Net architecture, the SE-Conformer model is also considered to explore the generalizability of the proposed modules. Specifically, the ABSF module is integrated into skip-connections and the CSAtt module is integrated to the end of each encoder layer and the beginning of each decoder layer, respectively. The ASR-oriented loss function is implemented by the multi-objective learning mechanism.

As shown in Table \ref{tab:se-conformer}, the number in the CSAtt column indicates the order of layers in which the CSAtt module is integrated into the encoder/decoder. For example, option 1 indicates that it is integrated only at the first layer of the encoder and the last layer of the decoder, and option 5 indicates that it is integrated at every layer of the encoder/decoder. The experimental results show that when ABSF, CSAtt, and $L_{ASR}$ are integrated into the SE-conformer, it achieves desired improvements on both SE and ASR tasks. In addition, the specific configurations of the CSAtt module are also determined by designed experiments. The results confirm the effectiveness of multi-objective learning for the task in this work with different model architectures. Most importantly, the results also prove the performance improvement of the ABSF, CSAtt and $L_{ASR}$ on both the CNN- and Conformer-based U-Net architecture, which confirm the contributions of all the proposed technical improvements in this work.

\section{Conclusions}
In this study, a recognition-oriented speech enhancement model (ROSE) based on multi-objective learning is proposed to achieve the SE task in the ATC domain, which enables us to improve the ASR performance without model retraining. Specifically, the ABSF and CSAtt modules highlight informative information and suppress redundant factors, enhancing the ability of the model to learn objective-related features. The ASR-oriented loss function is designed to guide the model to capture ASR-related representations in spectral scales. The multi-objective learning acquires the common representations across the two optimization objectives by combining the task-oriented loss function, achieving mutual promotion of SE and ASR tasks performance. The experiment results show that for both the SE and ASR tasks, the proposed framework can not only achieve the desired performance on the ATC corpus, but also be generalized to the public open-source datasets. The proposed framework can generate high-quality speech to address the speech echo in the ATC domain, as well as preferred ASR performance without any model retraining. In addition, all the proposed technical modules contribute to the expected performance improvements, as demonstrated by ablation studies. These improvements can also be applied to other neural architectures to enhance task performance. The visualization results showed that the proposed attention mechanism can learn informative weight distributions, providing interpretability for its performance effectiveness.

\bibliography{IEEE_bib_abbreviation}
\bibliographystyle{IEEEtran}

\end{document}